\documentclass[fleqn,usenatbib]{mnras}

\usepackage{newtxtext,newtxmath}

\usepackage[T1]{fontenc}

\DeclareRobustCommand{\VAN}[3]{#2}
\let\VANthebibliography\thebibliography
\def\thebibliography{\DeclareRobustCommand{\VAN}[3]{##3}\VANthebibliography}

\usepackage{graphicx}	
\usepackage{amsmath}	

\title[Polarized emission from 1E 2259+586]{The detection of polarized x-ray emission from the magnetar 1E 2259+586}

\author[Heyl et al.]{Jeremy Heyl$^{1}$,
Roberto Taverna$^{2}$,
Roberto Turolla$^{2,3}$,
Gian Luca Israel$^{4}$,
Mason Ng$^{5}$,
\newauthor
Demet Kirmizibayrak$^{1}$,
Denis Gonz\'alez-Caniulef$^{6}$,
Ilaria Caiazzo$^{7}$,
Silvia Zane$^{3}$,
Steven R. Ehlert$^{8}$
% ...
\\
The list of affiliations can be found at the end of the paper.
}

\date{Accepted XXX. Received YYY; in original form ZZZ}

\pubyear{2023}

\begin{document}
\label{firstpage}
\pagerange{\pageref{firstpage}--\pageref{lastpage}}
\maketitle

\begin{abstract}
We report on IXPE, NICER and XMM-Newton observations of the magnetar 1E~2259+586.  We find that the source is significantly polarized at about or above 20\% for all phases except for the secondary peak where it is more weakly polarized.  The polarization degree is strongest during the primary minimum which is also the phase where an absorption feature has been identified previously \citep{2019A&A...626A..39P}. The polarization angle of the photons are consistent with a rotating vector model with a mode switch between the primary minimum and the rest of the rotation of the neutron star.  We propose a scenario in which the emission at the source is weakly polarized (as in a condensed surface) and, as the radiation passes through a plasma arch, resonant cyclotron scattering off of protons produces the observed polarized radiation.   This confirms the magnetar nature of the source with a surface field greater than about $10^{15}$~G.
\end{abstract}




\section{Introduction}

\citet{1980Natur.287..805G} discovered the X-ray source 1E~2259+586 in observations of the supernova remnant G109.l$-$1.0 using the Einstein telescope on 17 December 1979.  Later identified as one of the  Anomalous X-ray Pulsars  \cite[AXPs;][]{mereste95}, 1E~2259+586 was the first member of this class to be discovered (and the third magnetar after SGR~1806$-$20 and SGR 0525$-$66 earlier in 1979).  \citet{1981Natur.293..202F} identified a periodicity in the X-ray emission from the source of 3.4890~s.  Subsequent observations revealed that the pulse profile exhibits two similar peaks over each rotation period.  Similar to the other anomalous X-ray pulsars the spin period of 1E~2259+586 is gradually increasing \citep[$\dot P \approx 5 \times 10^{-13}$~s~s$^{-1}$,][]{2014ApJ...784...37D}, and this is associated with the presence of a strong magnetic field ($\approx 6\times 10^{13}$~G) and the braking of the pulsar through magnetic dipole radiation.   Although the dipole magnetic field inferred from the spin down lies at the low end of the magnetar range \citep[in fact several radio pulsars have larger spin-down fields,][]{2005AJ....129.1993M}, 1E~2259+586 exhibits bursts, glitches and even an anti-glitch similar to other magnetars \citep{kasbelo17}, suggesting that a much stronger field of $10^{14}$--$10^{15}$~G, confined in small scale structures close to the surface, is present in 1E~2259+586, similarly to the other ``low-field'' magnetar SGR 0418+5729 \citep{2013Natur.500..312T}.

\citet{2019A&A...626A..39P} found evidence of phase-dependent spectral features in XMM-Newton observations of 1E~2259+586 that have been consistent in phase and energy over more than a decade while the source has been in quiescence.  On the other hand, during an active epoch in 2002, although the spectral feature was still present, its position in energy as a function of phase had changed.  \citet{2019A&A...626A..39P} proposed that this feature could be a signature of resonant cyclotron scattering similar to what has been proposed for a similar feature found in the phase-resolved spectrum of SGR 0418+5729 \citep{2013Natur.500..312T}.  If this association is correct in the case of 1E~2259+586, the  magnetic field for a proton cyclotron resonance is $(3-16)\times 10^{14}$~G (and below $10^{12}$~G for an electron cyclotron resonance).  Here we present polarimetric observations that provide evidence that this spectral feature results from photons scattering off of non-relativistic particles (most likely protons) at the cyclotron resonance.

\section{Observations}
\label{observations}

IXPE observed 1E 2259+5586 in June--July 2023 and a simultaneous XMM-Newton pointing was carried out on June 30 2023. NICER archival data, partially overlapping the IXPE time window, were also available. Tab.~\ref{tab:obslog} summarizes the observations used for this analysis.


\begin{table}
\begin{center}
    \caption{Observation Log.}\label{tab:obslog}
\begin{tabular}{rrr}
\hline
\multicolumn{1}{c}{Obs ID} & \multicolumn{1}{c}{MJD$_\textrm{Start}$} & \multicolumn{1}{c}{Exposure [s]}  \\
\hline
\multicolumn{3}{c}{IXPE} \\
02007899 & 60098 & 1202835 \\
\hline
\multicolumn{3}{c}{NICER} \\
6533040201  & 60022 & 358 \\ 
6533040301	& 60036	& 834 \\
6533040401	& 60050	& 1271 \\
6533040402	& 60064	& 1790 \\
6533040601	& 60078	& 698 \\
6533040701	& 60107	& 565 \\
6533040901	& 60120	& 541 \\
6533041001	& 60134	& 776 \\
6533041101	& 60149	& 564 \\
\hline
\multicolumn{3}{c}{XMM-Newton} \\
 0744800101 & 56868 & 112000 \\
0931790401  & 60126  & 20500 \\
\hline
\end{tabular}
\end{center}
\end{table}

\subsection{IXPE}
\label{obs_ixpe}


IXPE,  a NASA mission in partnership with the Italian space agency \citep[ASI;][and references therein]{2022JATIS...8b6002W} observed 1E~2259+5861 on 2023 June 2--19 and June 30--July 6 for 1.2~Ms in total.  The gas-pixel detectors on IXPE register the arrival time, sky position, and energy for each X-ray photon and use the photoelectric effect to provide an estimate of the position angle of each photon \citep{2021AJ....162..208S}. During each observation, photon arrival events registered between energies of $2$ and $8$ keV, within $48$ arcseconds of the position of source (R.A. $=345^\circ.3$, DEC $=58^\circ.9$), were extracted for analysis.  The background was estimated from an annular region centered on the source of inner and outer radii of $78$ arcseconds and $240$ arcseconds, respectively. 
Background subtraction was applied to the extracted Stokes parameters. Anyway, we observe that the background level for each IXPE detector unit (DU, i.e. telescope) in the $2$--$8$ keV band was less than $2\%$ of the source one. We could, hence, conclude that the energy-integrated analysis we performed is not much affected by background effects.
The times of the photon arrivals were finally corrected for the motion of IXPE around the barycenter of the Solar System.

\subsection{NICER}
\label{obs_nicer}
We used the Neutron star Interior Composition Explorer \citep[NICER;][]{2012SPIE.8443E..13G, 2016SPIE.9905E..4WL, 2016SPIE.9905E..1IP} data collected over 2023 March 19 to 2023 July 24. The data were processed with HEASoft version 6.31 and NICER Data Analysis Software (NICER-DAS) version 10 (2022-12-16\_V010a) using the \texttt{nicerl2} tool with standard filtering criteria, resulting in 6.1 ks of filtered exposure. We performed barycenter corrections in the ICRS reference frame using the JPL DE421 Solar System ephemeris with the \texttt{barycorr} tool in \texttt{FTOOLS} with coordinates ${\rm R.A.} = 345\fdg2845$, ${\rm DEC} = 58\fdg879$.


\subsection{XMM-Newton}
\label{obs_xmm}
A DDT pointing of 1E~2259+586 with XMM-Newton was activated on 2023, June 30, starting at 23:47:46 UTC for an exposure time of $\approx 20$ ks. The EPIC-pn \citep{str+01} as well as the two MOS cameras \citep{tur+01} were set in the Small Window mode, with a time resolution of $0.3$ s. Row data were processed by means of the \texttt{SAS} version 20.0 and the most updated calibration files.
After the subtraction of the intervals in which the background events were dominant, the data were extracted and processed applying standard procedures, for a net exposure of $\approx19.1$ ks for the MOSs and $\approx 18.8$ ks for the EPIC-pn. We extracted the source counts from a circular region of radius about 65". Those of the background were extracted from a similar region, within the same CCD where the source lies and $\sim$2.5' away for the pn, while for the MOSs, due to the use of the small window mode, from another CCD (at a distance of about 9' from the source). 
The times of the extracted photons were corrected for the barycenter of the Solar System in the ICRS reference frame with \texttt{barycen} tool in the \texttt{SAS}. The background subtracted source count rates were 10.71(3) ct s$^{-1}$ in the pn and 3.31(1) in the MOSs (1$\sigma$ confidence levels are reported). 



\section{Results}
\label{results}
\subsection{Timing analysis}\label{section:timing}

\begin{figure}
\begin{center}
\vspace{3mm}
\hspace{-5mm}\includegraphics[angle=-90,width=1.01\linewidth,clip,trim=0 0 0.25in 0]
{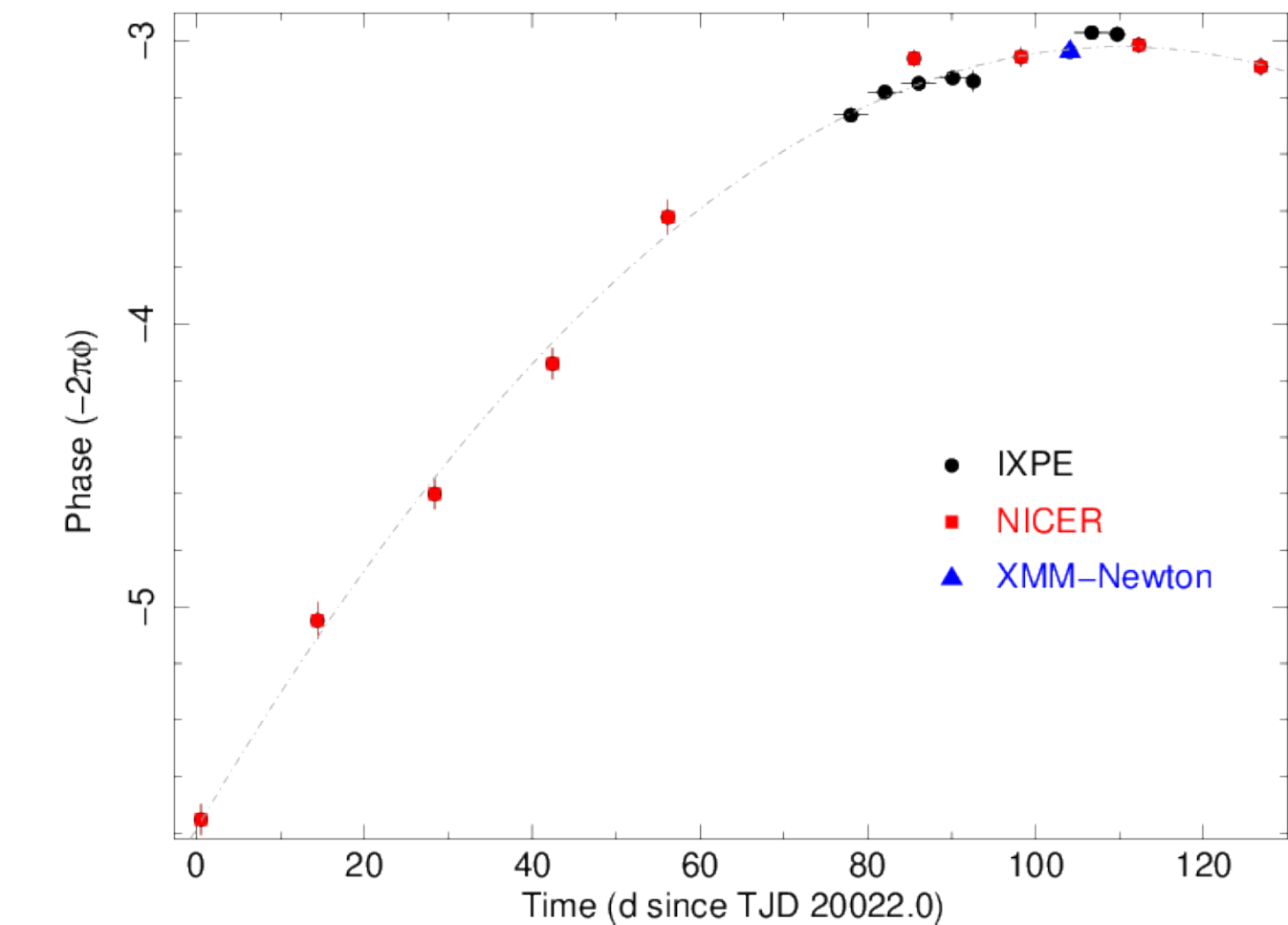} 
\phantom{MMM}Time [days since MJD 60022.0]
\end{center}
\caption{1E 2259+586 phase evolution (in radian units) as a function of time fitted with a linear plus a quadratic component for the whole data sample used in the analysis, that includes IXPE (black filled circles), NICER (red filled squares) and XMM-Newton (blue filled triangles). The dash-dotted line marks the best fit.}
\label{fig:phases}
\end{figure}
\begin{figure}
\centering\includegraphics[width=1.01\linewidth]{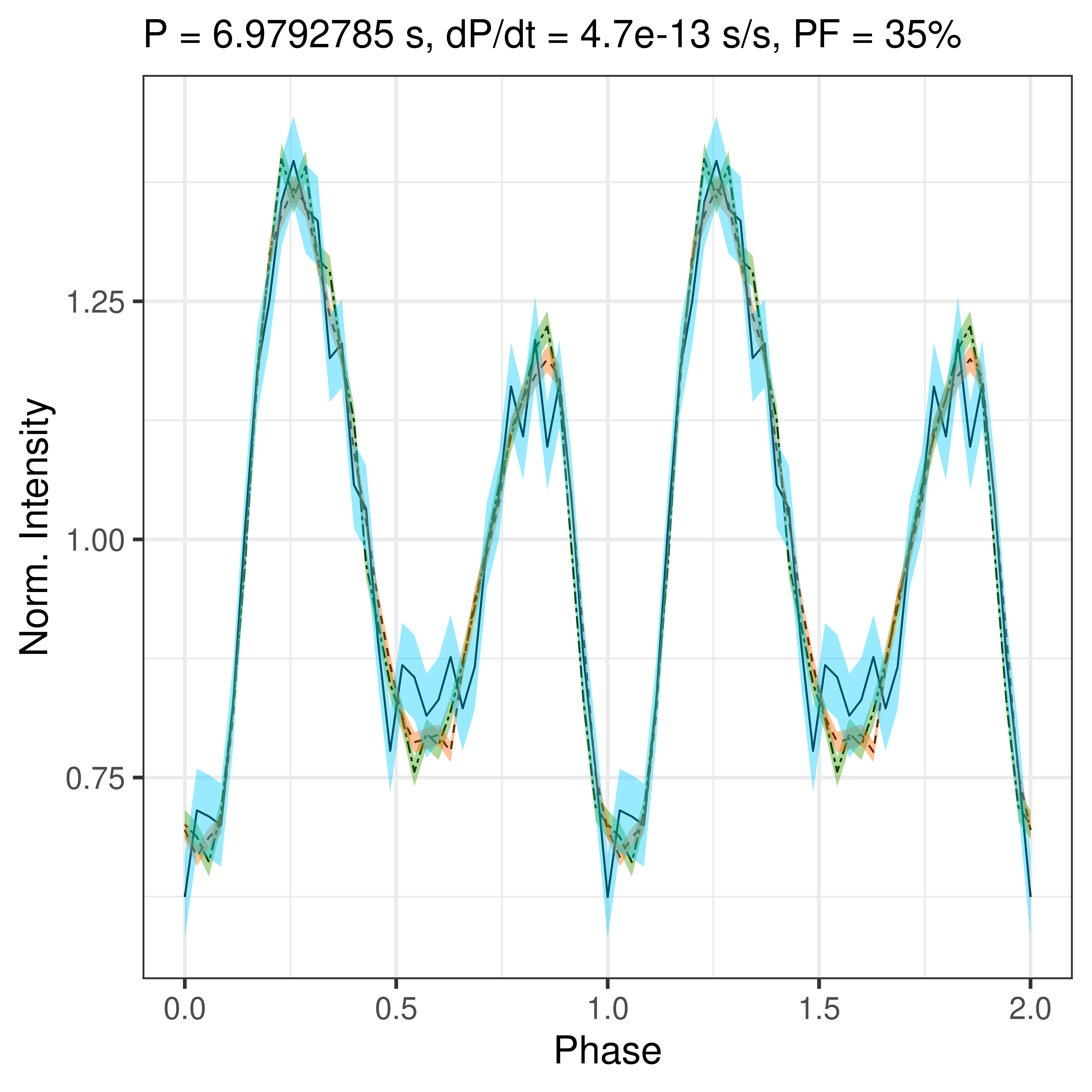}
\vspace{-7mm}
\caption{$1$--$10$\,keV NICER (cyan), $1$--$10$\,keV XMM-Newton (green) and $2$--$8$\,keV IXPE (orange) light curves folded to the best timing solution inferred from the whole sample of datasets and discussed in section\,\ref{section:timing} (see also Figure\,\ref{fig:phases}). }
\label{fig:FoldALL}
\end{figure}
\begin{figure}
\centering\includegraphics[width=0.98\linewidth]{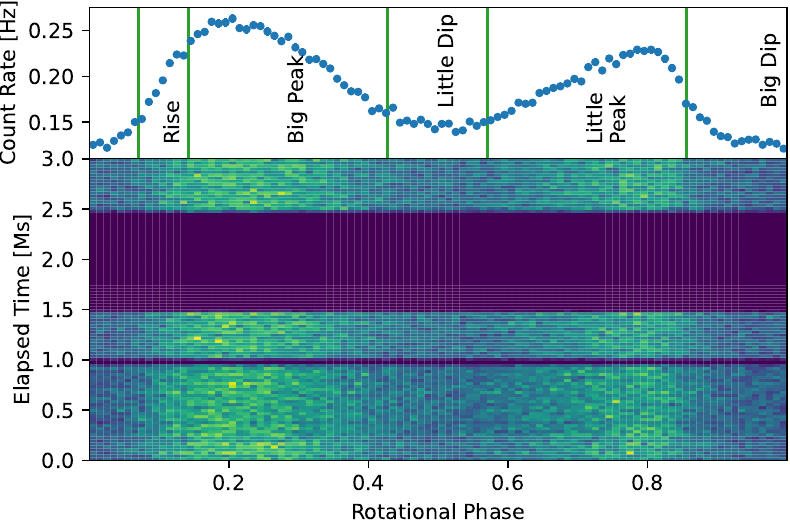}
\caption{IXPE Counts as a function of rotational phase and elapsed time for the inferred values of spin frequency and frequency derivative (see text for details). 
The zero phase and reference time are at MJD 60097.966874079. The spectral analysis phase regions are delineated and named in the upper panel (the Big Dip region spans across phase zero).}
\label{fig:folding}
\end{figure}

We first started analysing the datasets from each missions. In order to estimate the most reliable timing solution for the $1.2$ Ms exposure IXPE dataset, we used a phase-fitting approach  \cite[see, for example,][]{2003simone}  which gave a period of $6.979281(1)$ s, or $\nu = 0.14328124(3)$ Hz, at the reference epoch of 60097.0 MJD (a further $\dot P$ component did not significantly improve the fit). The peak-to-peak semi-amplitude of the background subtracted light curves folded to the above period resulted to be $(33\pm3)\%$. 
For NICER, the same phase-fitting algorithm revealed that a quadratic component was significantly present in the spin period phases as a function of time, resulting in the following timing solution: $P=6.9792783(1)$s and $\dot P = 5.0(2) \times 10^{-13}$ s s$^{-1}$, reference epoch 60022.0 MJD (1$\sigma$ confidence levels are reported), corresponding to $\nu = 0.143281290(3)$ Hz and $\dot \nu = -1.03(4) \times 10^{-14}$ Hz s$^{-1}$. Similarly, for XMM-Newton the best timing solution was inferred to be $P = 6.97931(2)$ s or $\nu = 0.1432806(4)$ Hz at reference epoch 60126.0 MJD. The inclusion of a first period derivative $\dot P$ did not significantly improve the fit. 
The peak-to-peak semi-amplitude of the background subtracted light curves folded to the above period resulted to be (35$\pm$3)\%. Note that the three timing solutions are in agreement with each other within their uncertainties.
Finally, the whole sample of IXPE, NICER and XMM-Newton datasets was used simultaneously to provide the best possible timing solution. A phase-fitting analysis (see Figure\,\ref{fig:phases}) gave the following result, $P=6.9792785(1)$s and $\dot P = 4.7(1) \times 10^{-13}$ s s$^{-1}$, reference epoch 60022.0 MJD, corresponding to $\nu = 0.143281286(2)$ Hz and $\dot \nu = -9.7(3) \times 10^{-15}$ Hz s$^{-1}$ (1$\sigma$ c.l.; reduced $\chi^2$$\sim$2 for 14 d.o.f. and rms of 0.007 cycles). 
In Figure\,\ref{fig:FoldALL} we show the light curves of each mission folded to the best solution discussed above.
The pulse shape is double peaked and does not change, within uncertainties, considering different energy bands. 
%
Figure\,\ref{fig:folding} 
shows the IXPE pulse profile with the different phase ranges used in the subsequent analysis: Big Dip, Rise, Big Peak, Little Dip and Little Peak (see Section \ref{pol_model} for further details); phase zero was chosen to coincide with that of \cite{2019A&A...626A..39P}. 

\subsection{Spectral analysis}
\label{sec:broadband}

\begin{figure*}
\centering\includegraphics[width=0.99\linewidth]{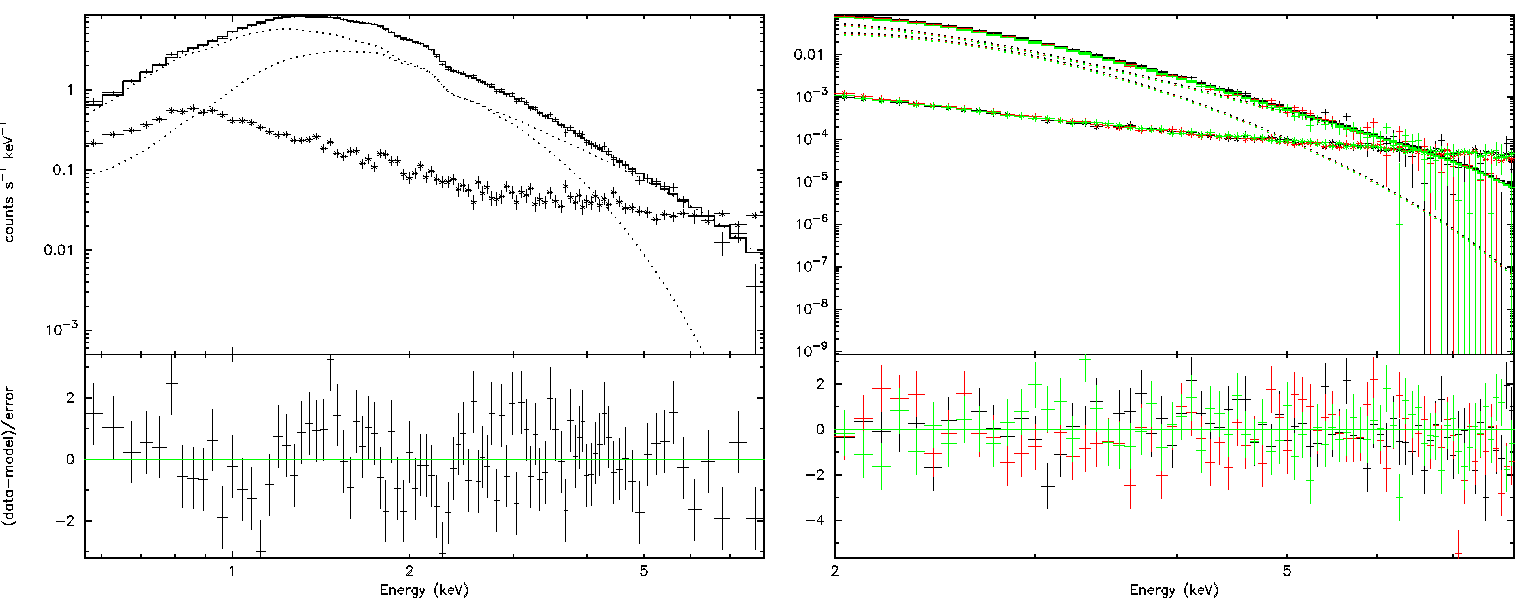}
\caption{Left: spectral fit of the EPIC-pn XMM-Newton data with the {\tt phabs}$\times${\tt (bbodyrad+powerlaw)}$\times${\tt gabs} model in the $0.5$--$8$\,keV range. Right: same for the IXPE DU1 (black), DU2 (red) and DU3 (green) data in the $2$--$8$\,keV range. The single spectral components are marked by dotted lines (see also Table\,\ref{tab:spec_ave}) and the background counts by crosses with error bars.}
\label{fig:specanalysis}
\end{figure*}

We first performed a phase-integrated spectral analysis of the XMM-Newton observation by fitting simultaneously the EPIC-pn and MOS data in the $0.5$--$8$ keV energy range using XSPEC \citep{Arnaud1996}. Fits with (absorbed\footnote{XSPEC model {\tt phabs}.}) single-component models (either a blackbody, BB, or a power-law, PL) turned out to be rather unsatisfactory, with a reduced $\chi^2$ exceeding $6$ for $250$ degrees of freedom (dof). A substantial improvement was obtained adding a second component. A purely thermal model (BB+BB), however, still resulted in a poor fit, with $\chi^2=462.2$ for 248 dof and a temperature for the hotter blackbody of $\sim 300$ keV, difficult to reconcile with the known properties of magnetars \cite[see e.g.][for reviews]{turrev15,kasbelo17}. By adopting a BB+PL decomposition the fit improved, although it was still far from being statistically acceptable,  
and the best-fitting parameters are compatible with those presented in previous works \cite[][]{zhu+08,2019A&A...626A..39P}. The addition of  a Gaussian absorption line, \cite[{\tt gabs} in XSPEC, as in][]{2019A&A...626A..39P}, resulted in a further improvement in the quality of the fit ($\chi^2=341.0$ for $245$ dof). By performing an f-test, the probability that the additional absorption line is unnecessary turns out to be $4.3\times10^{-7}$ (i.e. the feature is significant at $\sim5\sigma$ confidence level). The corresponding BB temperature, PL photon index, line energy and width are in agreement with those reported in \citet{2019A&A...626A..39P} within $1\sigma$ confidence level (see Table \ref{tab:spec_ave} for the fit parameters).  
The large  $\chi^2$ for the pn+MOS fit indicates that fitting the merged dataset to the phase-average spectrum may be inadequate.
Indeed, by restricting to the EPIC-pn data only resulted in a much better fit ($\chi^2=94.1$ for $93$ dof for the same spectral model, see Figure \ref{fig:specanalysis}), although the line centroid shifted to slightly higher energies. 
Results of the phase-resolved spectroscopic analysis are presented in Section \ref{spectropol_model}.

We finally attempted a fit of the IXPE $2$--$8$ keV data using the same spectral decomposition and freezing the column density and line parameters to those obtained from the fit of the EPIC-pn data. The fit is statistically acceptable ($\chi^2=138.0$ for 147 dof) with values of the free parameters in agreement (within statistical uncertainties) with those already obtained in the previous analyses (see again Table \ref{tab:spec_ave} and Figure \ref{fig:specanalysis}). 

\begin{table*}
\begin{center}
\caption{Results of the phase-integrated spectral fits with the model {\tt phabs}$\times${\tt (bbodyrad+powerlaw)}$\times${\tt gabs}.}
\label{tab:spec_ave}
\begin{tabular}{l|ccccccccc}
\hline
Data & $N_{\mathrm H}$ & $kT_{{\rm BB}}$ & $R_{{\rm BB}}\,^a$ & $\Gamma_{\rm PL}$ & Norm$_{\rm PL}$ at $1$ keV & $E_{\rm abs}$ & $\sigma_{\rm abs}$ & Depth$_{\rm abs}$ & $\chi^2$/dof \\
   & [$10^{22}\, \mathrm{cm}^{-2}$]  &  [keV]  &   [km]  &  & [$10^{-3}$/s/keV/cm$^2$] & [keV] & [keV] & [keV] & \\
\hline
PN+MOS & $0.91^{+0.08}_{-0.13}$ & $0.446^{+0.008}_{-0.009}$ & 2.24$^{+0.13}_{-0.13}$ & $3.93^{+0.08}_{-0.11}$ & $50.72^{+6.87}_{-9.14}$ & $0.71\pm^{+0.17}_{-0.22}$ & $0.30^{+0.09}_{-0.08}$ & $0.30^{+0.63}_{-0.20}$ & $341.0/245$ \\ 
\ & & & & & & & & & \\
PN & $1.02^{+0.03}_{-0.07}$ & $0.437^{+0.012}_{-0.011}$ & $2.33^{+0.22}_{-0.21}$ & $4.09^{+0.08}_{-0.08}$ & $62.09^{+6.81}_{-6.41}$ & $0.96^{+0.07}_{-0.18}$ & $0.23^{+0.10}_{-0.06}$ & $0.11^{+0.20}_{-0.05}$ & $94.1/93$ \\ 
\ & & & & & & & & & \\
IXPE$\,^b$ & $1.02\,^c$ & $0.429^{+0.011}_{-0.010}$ & $2.45^{+0.24}_{-0.20}$ & $4.36^{+0.09}_{-0.09}$ & $75.95^{+7.79}_{-7.80}$ & $0.96\,^c$ & $0.23\,^c$ & $0.11\,^c$ & $138.0/147$ \\ 

\hline
\hline
\end{tabular}
\begin{list}{}{}
\item[\,\,\,\,] Errors are at $1\sigma$ confidence level.
\item[$^a$] Derived by adopting a $3.2$ kpc distance \citep{kf12,2019A&A...626A..39P}.
\item[$^b$] For fitting IXPE data the spectral decomposition was convolved with a constant factor to take into account the different calibration of the 3 DUs \cite[the relative calibration factors obtained from the fit are compatible with those found in previous magnetar analyses, see][]{tav+22,zane+23,tur+23}.
\item[$^c$] Frozen to the value obtained from the PN fit.
\end{list}
\end{center}
\end{table*}

\subsection{Polarization analysis}
\label{sec:polarization}

\begin{figure}
\centering\includegraphics[width=0.98\linewidth]{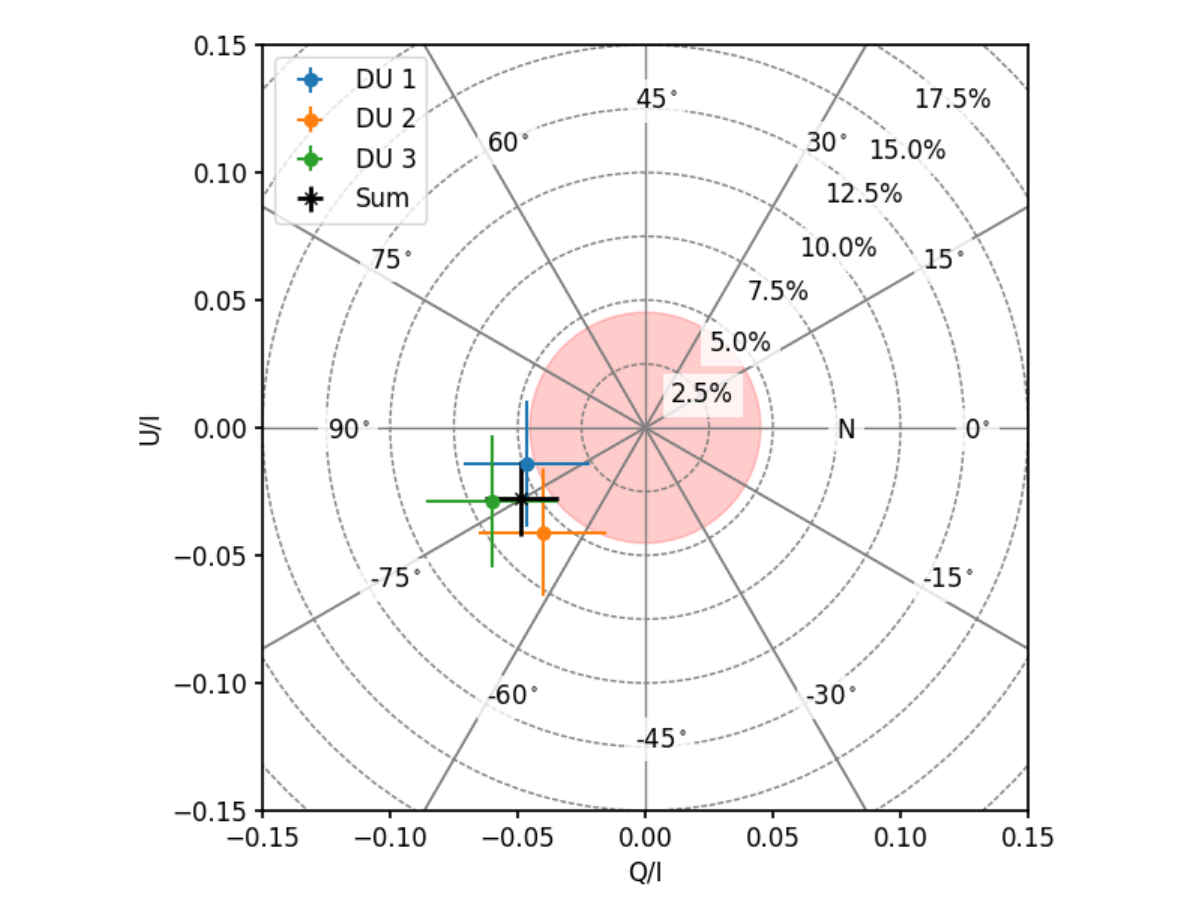}
\caption{Normalized Stokes parameters $Q/I$ and $U/I$ (filled circles with $1\sigma$ error bars) averaged over the rotational phase and integrated in the $2$--$8$ keV energy band for the single IXPE DUs (cyan, orange and green, respectively) for the sum (black). The gray-dotted circles represents the levels of constant PD, while the gray-solid lines those of constant PA (measured East of North). The red shaded region represents the MDP at $99\%$ confidence level for the combined measurement.}
\label{fig:ph-en-int}
\end{figure}

A phase- and energy-integrated study (in the $2$--$8$ keV band) of the polarization properties of the source was performed using the {\tt PCUBE} algorithm of the {\sc ixpeobssim} suite \cite[][]{BALDINI2022101194}\footnote{\url{https://github.com/lucabaldini/ixpeobssim}}. The results for the normalized Stokes parameters $Q/I$ and $U/I$ are reported in Figure \ref{fig:ph-en-int} for the single IXPE DUs and for the sum of them. In the figure the loci of constant polarization degree (PD $=\sqrt{(Q/I)^2+(U/I)^2}$) and polarization angle (PA $=\arctan(U/Q)/2$) are also shown, together with the value of the minimum detectable polarization at $99\%$ confidence level \cite[MDP$_{99}$;][]{weiss10}. We obtained a highly probable detection (significance $>99.9\%$), with PD $=5.6\pm1.4\%$ (above the MDP$_{99}=4.5\%$) and PA $=-75^\circ.2\pm7^\circ.4$ measured East of North (errors at $1\sigma$).

We performed  a phase-integrated, energy-dependent polarimetric analysis as well, by dividing the $2$--$8$\,keV band into 3 bins. The only bin with a polarization degree in excess of the MDP$_{99}$ is that at low energies ($2.0$--$3.2$\,keV), with PD $=6.1\pm1.5\%$ (MDP$_{99}=4.6\%$) and PA $=-66^\circ.4\pm7^\circ.1$ (errors at $1\sigma$). In the rest of the energy range we can derive only upper limits: PD $<14.6\%$ in the $3.2$--$5.0$\,keV range and PD $<70.0\%$ in the $5.0$--$8.0$\,keV one, at $3\sigma$ confidence level. However, the null hypothesis probability that no polarization is detected is below $3\times 10^{-4}$.

 In order to study the evolution of the polarization with the rotational phase, we divided the counts into 14 equally spaced phase bins taking as phase zero the one reported in  \citet{2019A&A...626A..39P}.  In each bin we calculate the normalized Stokes parameters $Q/I$ and $U/I$ using the likelihood method outlined in \citet{gonzalezcaniulef_unbinned}; similar results can be achieved using the standard tools (e.g. {\sc ixpeobssim}).  The results are given in Table~\ref{tab:polarization}. There (as well as in the central panel of Figure~\ref{figure:bestfitRVM}) we listed the values of polarization degree and polarization angle even for the bins where the PD lies below the $\textrm{MDP}_{99}$. This is justified because we measured an overall polarization with high significance (see above), so that the null hypothesis relevant here is not that of unpolarized radiation, but rather of constant polarization. This allows us to probe the agreement between the binned results and the polarization models which are fit directly to the data for individual photons \citep{gonzalezcaniulef_unbinned}.  
 
\begin{figure}
    \centering
    \includegraphics[width=0.98\linewidth]{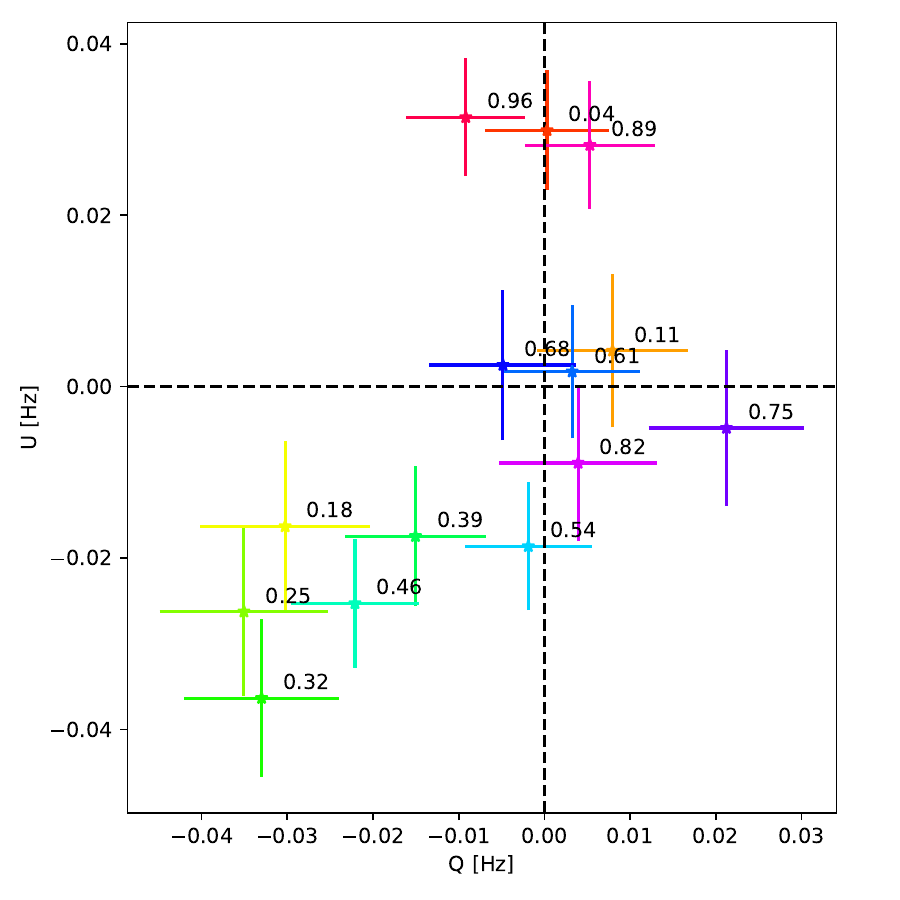}
    \caption{The Stokes Parameters $Q$ and $U$ as a function of phase in units of the mean number of counts per second observed with IXPE in the $2$--$8$~keV range. The uncertainties correspond to $\Delta \log L=1/2$ contours of the likelihood. Each cross is labeled by the central value of the corresponding phase bin and coloured according to phase starting with red at phase zero and proceeding along the colour wheel through green then blue.}
    \label{fig:QU-phase}
\end{figure}

 The Stokes parameters for the bins reported in Table~\ref{tab:polarization} are also shown in Fig.~\ref{fig:QU-phase}.
 There, three distinct regimes in phase can be recognized. The polarized flux is maximized during the Big Dip (the cluster at the top of Figure~\ref{fig:QU-phase}) and the Big Peak (the cluster at the bottom left).  The polarized flux during the Little Dip is modest, and the polarized flux in the Little Peak and the Rise is very low.

\begin{table*}
    \centering
    \caption{The normalized Stokes parameters, polarization degree and angle (2$-$8~keV) for the IXPE observation of 1E~2259+5861 in the different phase bins. The uncertainties correspond to $\Delta \log L=1/2$ contours of the likelihood \citep{gonzalezcaniulef_unbinned}.}
    \label{tab:polarization}
    \begin{tabular}{cr|rrrrrrc}
    \hline \hline
    N &
\multicolumn{1}{c}{Phase Range} &        
\multicolumn{1}{c}{$Q/I$} & 
\multicolumn{1}{c}{$U/I$} &
\multicolumn{1}{c}{$\textrm{MDP}_{99}$} &
\multicolumn{1}{c}{PD} &
\multicolumn{1}{c}{PA [deg]} &
\multicolumn{1}{c}{Counts} & 
\multicolumn{1}{c}{Region}\\ \hline \hline
 1 & $0.000-0.071$ &  $ 0.002\pm0.054$  &   $ 0.225\pm0.052$  & 0.163 &   $ 0.225\pm0.051$  &   $ 44.7\pm~~6.9$  &    11443 & Big Dip \\
 2 & $0.071-0.143$ &  $ 0.040\pm0.045$  &   $ 0.021\pm0.045$  & 0.135 &   $ 0.045\pm0.041$  &   $ 13.9\pm 27.7$  &    16962 & Rise    \\
 3 & $0.143-0.214$ &  $-0.119\pm0.039$  &   $-0.064\pm0.039$  & 0.118 &   $ 0.136\pm0.036$  &   $-75.8\pm~~8.1$  &    21788 & Big Peak\\
 4 & $0.214-0.286$ &  $-0.140\pm0.039$  &   $-0.105\pm0.039$  & 0.118 &   $ 0.175\pm0.037$  &   $-71.6\pm~~6.3$  &    21445 & Big Peak \\
 5 & $0.286-0.357$ &  $-0.151\pm0.041$  &   $-0.167\pm0.042$  & 0.126 &   $ 0.225\pm0.039$  &   $-66.1\pm~~4.9$  &    18739 & Big Peak\\
 6 & $0.357-0.429$ &  $-0.086\pm0.047$  &   $-0.099\pm0.046$  & 0.141 &   $ 0.131\pm0.044$  &   $-65.4\pm~~9.8$  &    15111 & Big Peak \\
 7 & $0.429-0.500$ &  $-0.146\pm0.049$  &   $-0.167\pm0.050$  & 0.150 &   $ 0.222\pm0.050$  &   $-65.6\pm~~6.1$  &    12988 & Little Dip\\
 8 & $0.500-0.571$ &  $-0.013\pm0.050$  &   $-0.127\pm0.051$  & 0.153 &   $ 0.128\pm0.049$  &   $-47.9\pm 10.5$  &    12589 & Little Dip\\
 9 & $0.571-0.643$ &  $ 0.020\pm0.049$  &   $ 0.011\pm0.048$  & 0.145 &   $ 0.023\pm0.050$  &   $ 14.2\pm 57.0$  &    14039 & Little Peak\\
10 & $0.643-0.714$ &  $-0.025\pm0.045$  &   $ 0.013\pm0.046$  & 0.136 &   $ 0.029\pm0.045$  &   $ 76.3\pm 32.5$  &    16426 & Little Peak\\
11 & $0.714-0.786$ &  $ 0.098\pm0.042$  &   $-0.022\pm0.042$  & 0.126 &   $ 0.100\pm0.041$  &   $ -6.4\pm 11.4$  &    18709 & Little Peak\\
12 & $0.786-0.857$ &  $ 0.018\pm0.043$  &   $-0.042\pm0.043$  & 0.130 &   $ 0.045\pm0.042$  &   $-33.1\pm 26.5$  &    18389 & Little Peak\\
13 & $0.857-0.929$ &  $ 0.036\pm0.052$  &   $ 0.194\pm0.051$  & 0.156 &   $ 0.197\pm0.050$  &   $ 39.7\pm~~7.2$  &    12495 & Big Dip\\
14 & $0.929-1.000$ &  $-0.072\pm0.055$  &   $ 0.247\pm0.054$  & 0.165 &   $ 0.257\pm0.053$  &   $ 53.1\pm~~5.8$  &    10920 & Big Dip \\

\hline
    \end{tabular}
\end{table*}

\section{Polarization Modeling}
\label{pol_model}
\begin{figure}
\begin{center}
\includegraphics[width=0.98\linewidth]{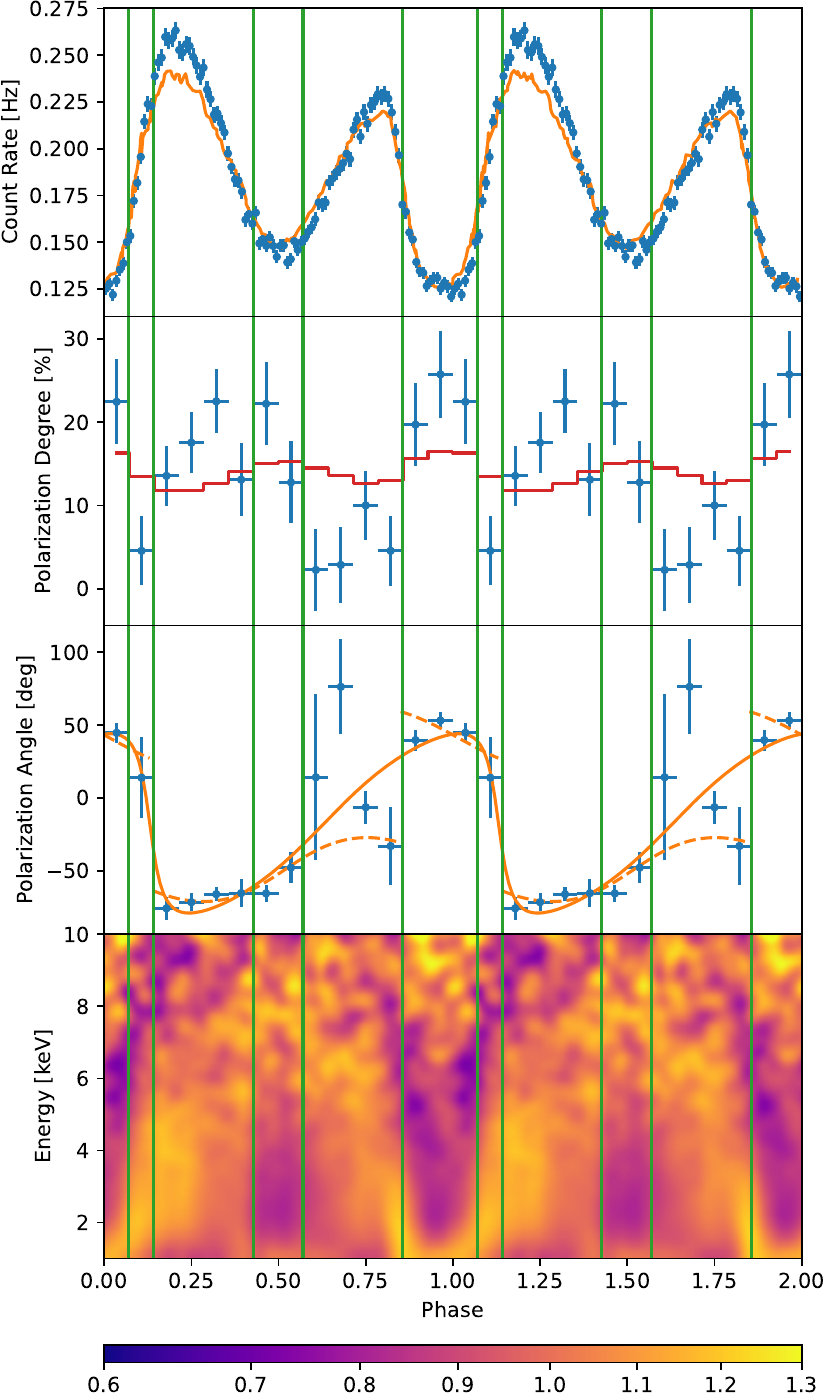}
\caption{IXPE and XMM-Newton energy-integrated ($2$--$8$ keV range) counts (upper), PD (upper-center) and PA (lower and lower-center, blue points with error bars). 
The uncertainties in the lower panels correspond to $\Delta \log L=1/2$ contours of the  likelihood.  The orange curve in the upper panel is the XMM-Newton ($0.3$--$12$~keV) pulse profile from \citet{2019A&A...626A..39P} scaled to the mean count rate of IXPE.  The red curve  in the second panel depicts the value of MDP$_{99}$ for each bin. 
The orange curves in the third panel show the best-fitting RVM model without (solid) or with (dashed) mode switching
The lowermost panel shows the phase-resolved spectrum observed in 2014 with XMM-Newton EPIC-pn \citep[as in Fig. 2 of][]{2019A&A...626A..39P}, normalized to the phase-averaged energy spectrum and energy-integrated pulse profile. The vertical green lines mark the five phase intervals used in our analysis.
}
\label{figure:bestfitRVM}
\end{center}
\end{figure}


In the highly magnetized environment surrounding a magnetar radiation propagates into two (normal) modes, the ordinary (O) and extraordinary (X) ones. In the former case, the electric field of the wave oscillates in the plane of the (local) magnetic field and of the photon momentum, while in the latter the oscillations are perpendicular to this plane \citep{1978SvAL....4..117G,1979JETP...49..741P}. Even in the absence of matter, vacuum birefringence will force the polarization vector of photons to follow the direction of the (local) magnetic field until the so-called polarization-limiting radius \citep{2000MNRAS.311..555H,2002PhRvD..66b3002H}. 
For typical magnetars, this radius is
estimated to be about $200$--$300$ stellar radii for keV photons \cite[see, e.g., Figure 1 of][and also \citealt{2018Galax...6...76H}]{taverna+2015}, where the field is dominated by the dipole component. The polarization measured at the telescope is, then, expected to be either parallel or perpendicular to the instantaneous projection of the magnetic dipole axis of the star onto the plane of the sky. For this reason, the modulation of the polarization angle with phase is decoupled from the evolution of the polarization degree and intensity
(that carry the imprint of the conditions at emission) and most likely should follow the rotating vector model \citep[RVM;][see also \citealt{tav+22,gonzalezcaniulef_unbinned}]{1969ApL.....3..225R,2020A&A...641A.166P},
\begin{equation}
    \tan(\textrm{PA}-\chi_{\rm p}) = \frac{\sin\theta\sin(\phi-\phi_0)}{\cos i_{\rm p} \sin\theta\cos(\phi-\phi_0)-\sin i_{\rm p} \cos\theta } ,
\end{equation}
where $i_{\rm p}$ is the inclination of the spin axis with respect to the line-of-sight, $\chi_{\rm p}$ the position angle of the spin axis in the plane of the sky (measured East of North), $\theta$ the inclination of the magnetic axis to the spin axis and $\phi$ is the spin phase ($\phi_0$ is the initial phase).  The angle between the dipole axis and the line-of-sight varies between $i_{\rm p}-\theta$ at $\phi=\phi_0$ and $i_{\rm p}+\theta$ at $\phi=\phi_0+\pi$.  Without loss of generality we restrict the parameters as follows:
$$ 
0\leq i_{\rm p}\leq 180^\circ, 0\leq \theta \leq 90^\circ, 0\leq \chi_{\rm p} < 180^\circ, 0\leq\phi_0<360^\circ.
$$

\begin{figure*}
    \includegraphics[width=0.9\textwidth]{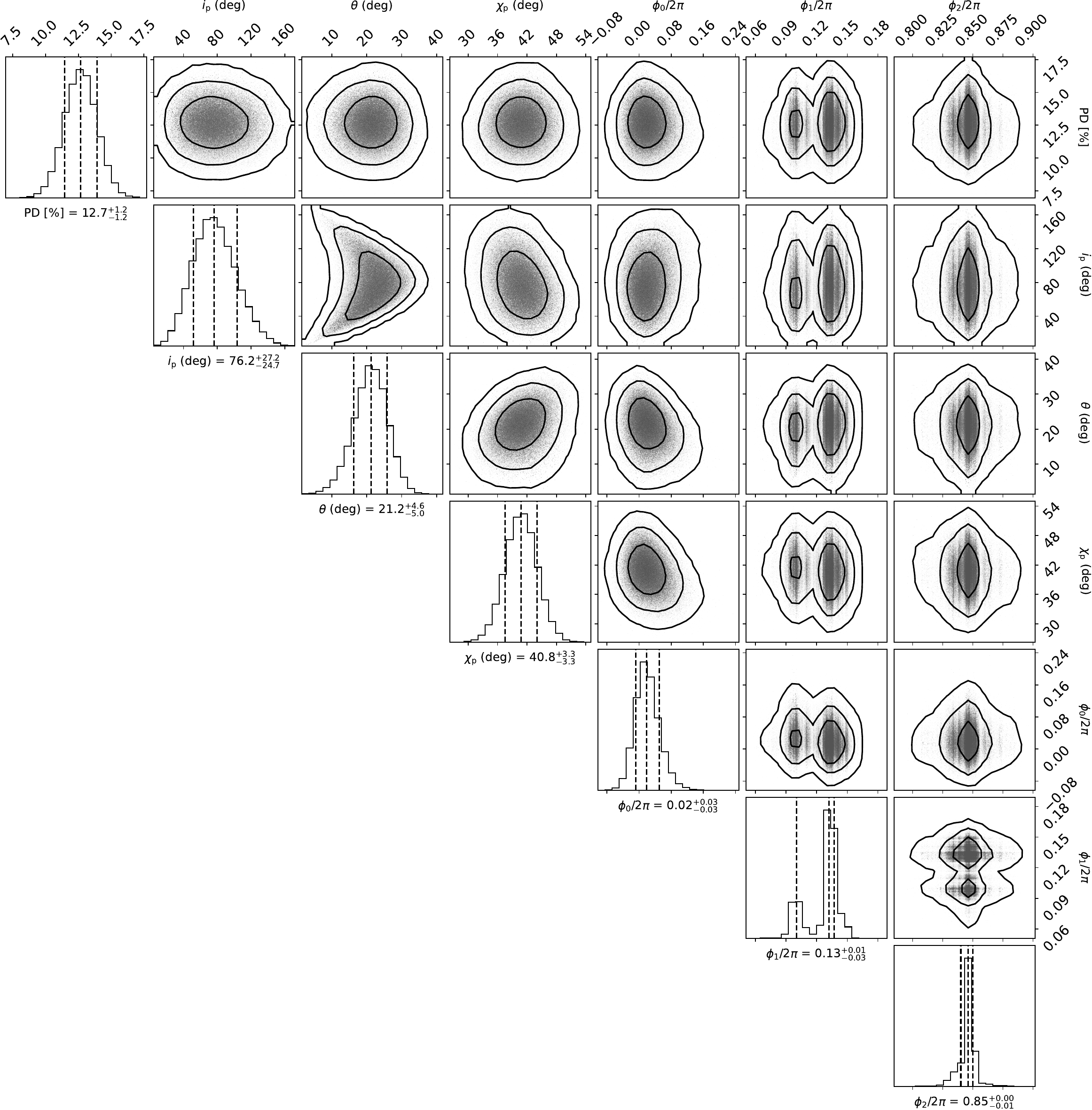}
    
    \vspace{-0.62\textwidth}
    \hspace{-0.34\textwidth}
    \includegraphics[width=0.65\textwidth]{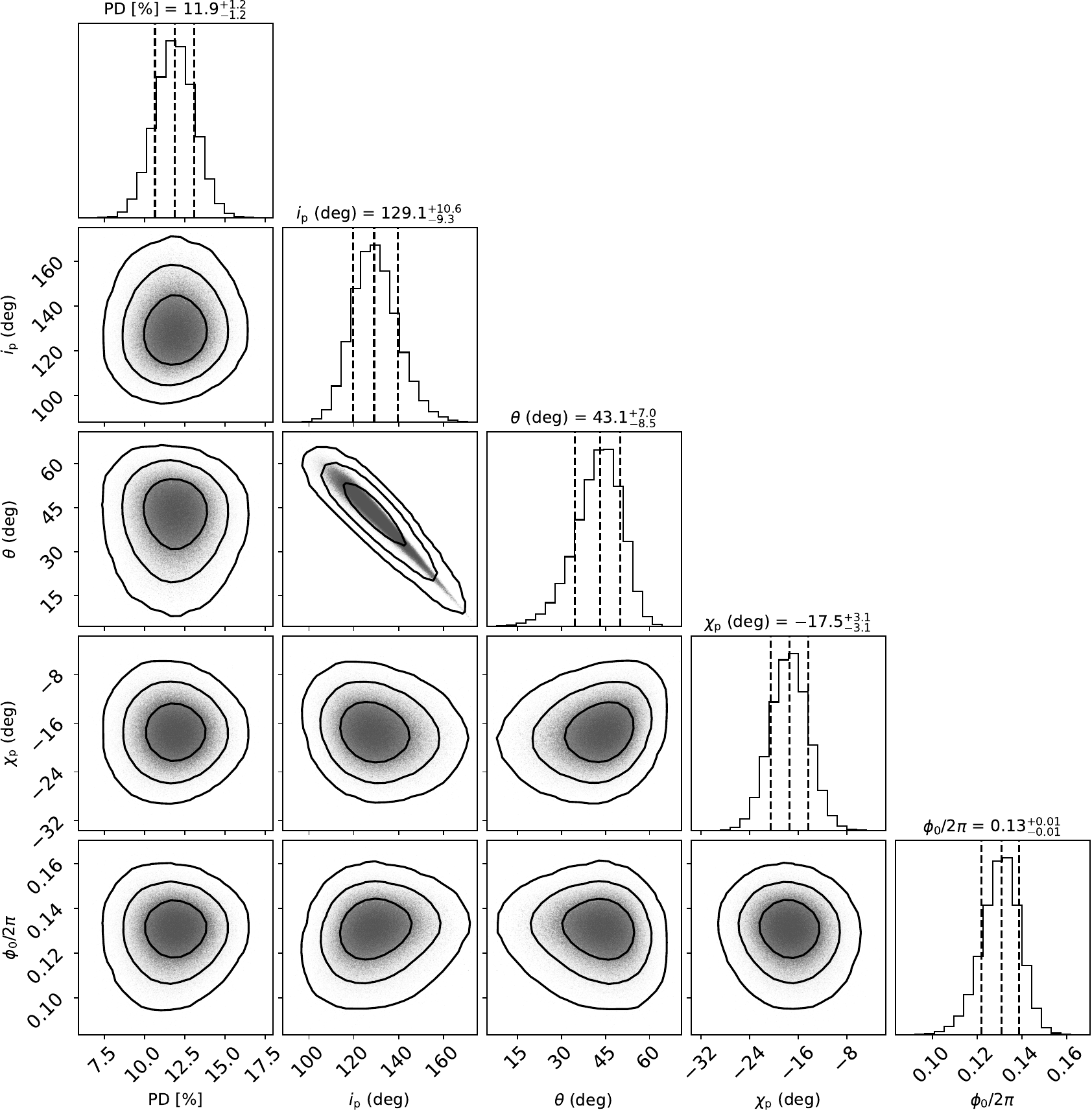}
    \caption{Posteriors of the RVM for the IXPE observations of 1E 2259+5586 The two-dimensional contours correspond to $68\%$, $95\%$ and $99\%$ confidence levels.  
The histograms show the normalized one-dimensional distributions for a given parameter derived from the posterior samples. The upper grid depicts constraints on the RVM with a mode switch between phases $\phi_1$ and $\phi_2$.  The lower grid depicts the constraints on the RVM without a mode switch.} \label{fig:geometry}
\end{figure*}

The two best-fitting RVMs for the polarization angle are depicted in Figure~\ref{figure:bestfitRVM}.  The solid curve traces a model where the polarization mode is constant with phase, and the dashed curve shows a model where the polarization mode switches at a phase where the polarization degree is low.  This is accomplished by replacing the Stokes parameters ($Q$ and $U$) of the model by their additive inverse over a range of phases $\phi_1 < \phi < \phi_2$ where $\phi_1$ and $\phi_2$ are two additional parameters.  The log-likelihood of the first model (with $\textrm{PD}=11.6\%$) is 49.6 and that of the second one ($\textrm{PD}=12.7\%$) is 57.8. 
The log-likelihoods for 222,043 events, drawn from two models with the observed values of  PDs, are
distributed approximately normally, with means of $48$ and $56$ and standard deviations of $10$; this indicates that both models are good fits to the data; about 60\% of the time random events drawn from these models will yield likelihoods smaller than those measured for the data.  Figure~\ref{fig:geometry} depicts posterior distributions of the parameters for these two models.

\begin{table*}
\centering
\caption{Best-fitting RVM parameters. When the data are consistent with the model, the log-likelihood ($\log L$) is normally distributed. The fit quality in the last column is given in terms of the best-fitting log-likelihood compared with the expected value, where positive values indicate better-than-expected values.} \label{tab:rvm_param}
\begin{tabular}{l|cccccccc}
\hline
\hline
& 
Mean PD &  $\chi_{\rm p}$ &  $\theta$ & $i_{\rm p}$ & $\phi_0/2\pi$  & $\phi_1/2\pi$ & $\phi_2/2\pi$ & $\Delta\log L$ \\
&  [\%] & [deg] &   [deg] &  [deg] &  &  & & [$\sigma$] \\ 
\hline  
\hline  
Mode-Switching Model  & $ 12.7 \pm 1.2$ & $\!~~~76.2^{+27.2}_{-24.7}$ & $21.2^{+4.6}_{-5.0}$ & $~~~40.8\pm3.3$ & $0.02\pm0.02$  & $0.13^{+0.01}_{-0.03}$ & $0.85\pm0.01$ & $+0.18$ \smallskip \\
Single-Mode Model & $ 11.9 \pm 1.2$ & $129.1^{+10.6}_{-9.3}$ & $43.1^{+7.0}_{-8.5}$ & $-17.5\pm3.1$ & $0.13\pm0.01$  & --- & ---- & $\!+0.16$ \\
\hline
\end{tabular}
\end{table*}

\begin{figure}
    \centering
    \includegraphics[width=0.9\linewidth]{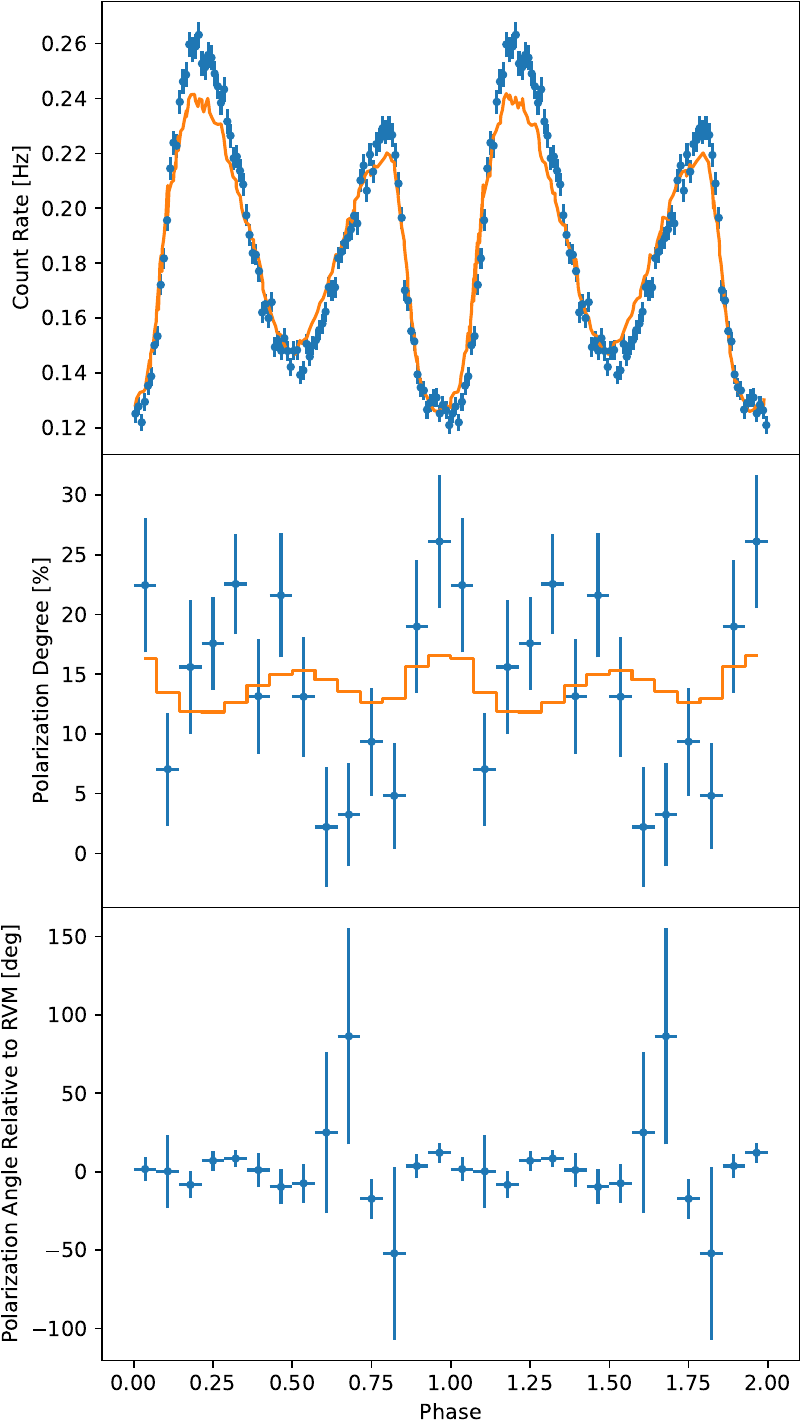}
    \caption{Same as in the upper, upper-center and lower-center panels of Figure~\ref{figure:bestfitRVM}, but with Stokes parameters referred to the frame of the best-fitting RVM without mode switching.
    The polarization angles are generally consistent with zero within the uncertainties, while the polarization degree is similar to that shown in Figure~\ref{figure:bestfitRVM}, except for phase 0.1, during the rapid swing in the RVM, where the polarization is somewhat washed out in Figure~\ref{figure:bestfitRVM}.  However, even when correcting for the rotation, the polarization degree at this phase lies below MDP$_{99}$.}
    \label{fig:triple_rot}
\end{figure}

As the model without mode switching is nested within the mode-switching model, we can calculate the probability to achieve the measured likelihood ratio even if there is no mode switch (the null hypothesis).  Twice the difference in likelihoods is distributed as a $\chi^2$ distribution with two degrees of freedom (for the two additional parameters), yielding a probabilty that the null hypothesis is true of less than  $3\times 10^{-4}$.   Furthermore, one of the mode switches can account for the low polarization at phase 0.11 that lies between two high polarization regimes.   Figure~\ref{fig:triple_rot} examines the single-mode RVM in more detail by removing the effect of the motion of the magnetic axis on the plane of the sky from the measured photon polarization angles. To this aim, the $Q$ and $U$ Stokes parameters for each photon are rotated into the frame of the best-fitting RVM before measuring the polarization angle and degree. If the low polarization at phase 0.11 results from the smearing of a large intrinsic polarization as the star rotates, the polarization measured after this procedure would be large. However, as Figure~\ref{fig:triple_rot} shows, the polarization at this phase remains low, so it is indeed a natural time for a mode switch as found in the mode-switching model indicated by dashed lines in Figure~\ref{figure:bestfitRVM}.  Both the  single-mode and the mode-switching models allow for solutions with $i_{\rm p}>90^\circ$ such that the dipole axis points closest toward the line-of-sight \citep[and so the emission is expected to be brighter,][]{1998MNRAS.300..599H} at phase 0.13 and 0.02, respectively, landing just before the Big Peak of the light curve.  For $i_{\rm p}<90^\circ$, a secondary peak lies at about phase zero.  However, it is obvious from Figure~\ref{figure:bestfitRVM} that phase zero is the deepest minimum in the light curve. Beyond the mode switching itself, a key difference between the models is that the model without mode switching requires the angle between the magnetic axis and the spin axis ($\theta$) to be larger ($43^\circ$) than what is expected for the mode-switching model ($21^\circ$), in order to account for the observed swing in polarization angle between the Big Dip and the Big Peak.  In the mode switching model, this is accomplished with a smaller magnetic obliquity plus a mode switch at phases 0.1 and 0.85, so during the Big Dip the emission is dominated by a different polarization mode than the rest of the time.  

In the mode-switching model the emission should peak right in the middle of the Big Dip, if it is associated with the orientation of the dipole field. 
If one ignores the phase region of the Big Dip, the rest of the pulse profile can result from a single hot spot about $10^\circ$ in radius located about $20^\circ$ from the spin axis, and with the spin axis pointing about $100^\circ$ from the line-of-sight (as in Figure~\ref{fig:geometry}).  These considerations, along with the unique polarization signature of the Big Dip, point toward the hypothesis that the basic emission geometry of the pulsar is straightforward with some sort of obscuration that operates during the Big Dip. 

\begin{figure}
    \begin{center}
        \includegraphics[width=0.97\linewidth]{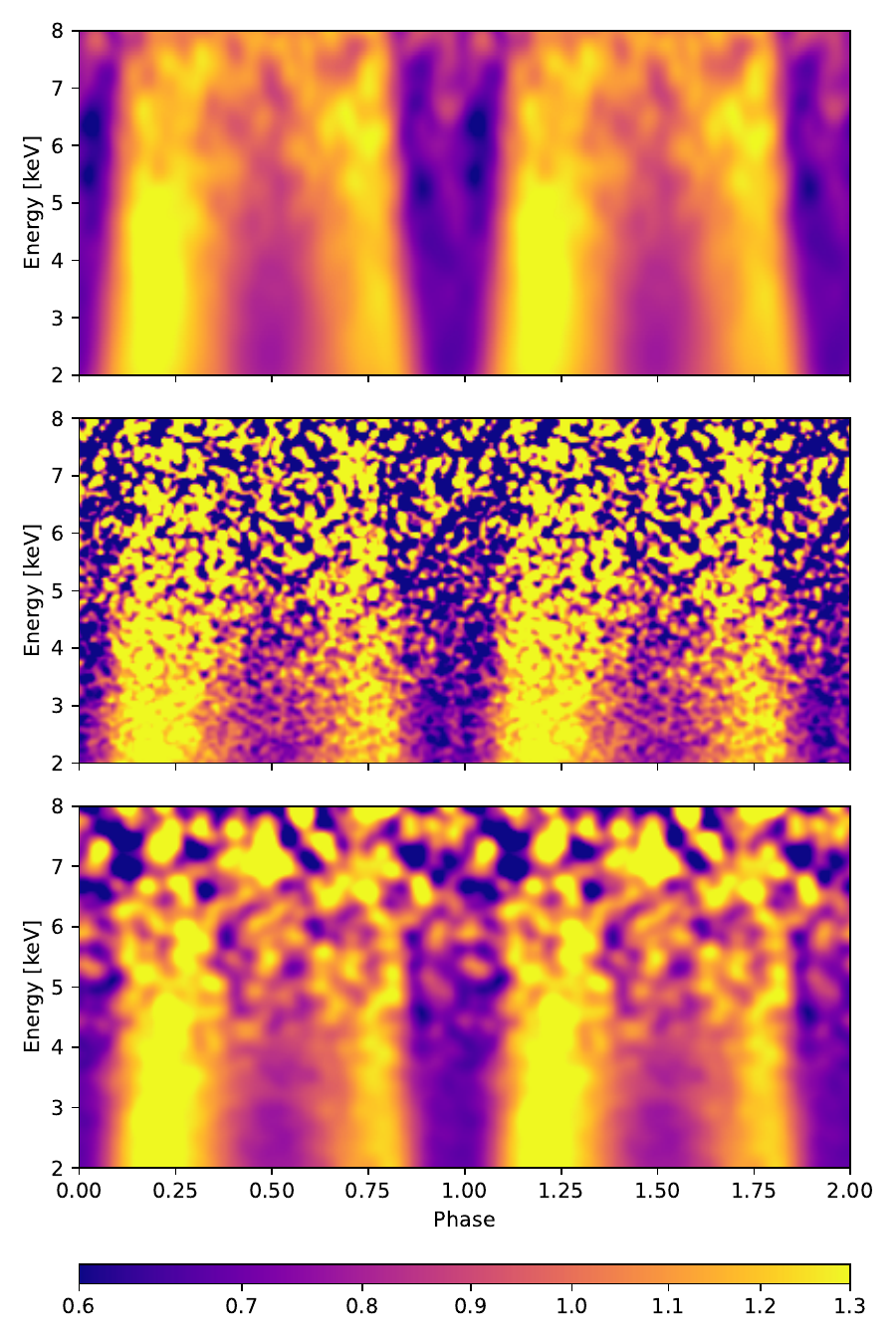}
        \caption{Phase-resolved spectra of 1E~2259+586 in the $2$--$8$~keV range as observed by XMM-Newton EPIC-pn in 2014 
        \citep[upper panel, as in Fig. 3 of][]{2019A&A...626A..39P}, by XMM-Newton EPIC-pn and MOS in 2023 (middle panel), and by IXPE in 2023 (lower panel).
        These are normalized to the phase-averaged energy spectrum but not normalized to the energy-integrated pulse profile.
        }
        \label{fig:dynamic3}
    \end{center}
\end{figure}

\citet{2019A&A...626A..39P} found evidence for a spectral feature around phase zero (the Big Dip) that appears consistently in XMM-Newton observations of 1E~2259+586 in quiescence in 2002 and again in 2014.  When the source was in outburst in 2002, a feature appears but with a different phase dependence.  The pulse profile that we have observed with IXPE is consistent with that observed with XMM-Newton in 2014, as shown in the upper panel of Figure~\ref{figure:bestfitRVM}.  The lowermost panel of Figure~\ref{figure:bestfitRVM} depicts the phase-resolved spectrum of 1E~2259+586 in 2014, indicating the presence of the spectral feature coincident with the large dip in the pulse profile of the source. This is further highlighted in Figure~\ref{fig:dynamic3}, where the $2$--$8$~keV, phase-resolved spectra for the XMM-Newton 2014 and 2023 observations are shown together with the IXPE one.
Although the signal-to-noise ratio of the shorter 2023 observations is lower than that of the 2014 ones, the absorption feature is discernible both in the IXPE and the latest XMM-Newton observations as well. 

\citet{2019A&A...626A..39P} argue that this feature may be due to resonant cyclotron scattering of X-rays off of protons in the magnetosphere. In analogy to the results of  \citet{2013Natur.500..312T}, the scattering plasma should be confined along a magnetic loop above the surface, with proton cyclotron energy ranging from about $2$ to $10$ keV, corresponding to a magnetic field strength along the loop from $3\times 10^{14}$~G to $2\times 10^{15}$~G, neglecting gravitational redshift. 

To examine the implications of this scenario on the polarized flux from the surface of the neutron star, we adapt the expressions for electron resonant cyclotron scattering cross sections in the non-relativistic limit \cite[][]{2008MNRAS.386.1527N} to the case of scattering off protons,
\begin{equation} \label{eqn:RCScrosssect}
\begin{array}{lll}
\dfrac{\textrm{d}\sigma}{\textrm{d}\Omega}_{\textrm{O}\rightarrow \textrm{O}} &=&  \dfrac{3\pi r_0 c}{8} \delta(\omega-\omega_B) \cos^2\beta \cos^2 \beta' \\ 
&& \\
\dfrac{\textrm{d}\sigma}{\textrm{d}\Omega}_{\textrm{O}\rightarrow \textrm{X}} &=&  \dfrac{3\pi r_0 c}{8} \delta(\omega-\omega_B) \cos^2\beta  \\ 
&& \\
\dfrac{\textrm{d}\sigma}{\textrm{d}\Omega}_{\textrm{X}\rightarrow \textrm{X}} &=&  \dfrac{3\pi r_0 c}{8} \delta(\omega-\omega_B)  \\ 
&& \\
\dfrac{\textrm{d}\sigma}{\textrm{d}\Omega}_{\textrm{X}\rightarrow \textrm{O}} &=&  \dfrac{3\pi r_0 c}{8} \delta(\omega-\omega_B) \cos^2 \beta'\,,
\end{array}
\end{equation}
where $\beta$ ($\beta'$) is the photon angle with respect to the magnetic field before (after) scattering, $\omega$ is the photon frequency, $\omega_B$ is the proton cyclotron frequency and $r_0$ is the classical proton radius (a factor of 1,836 smaller than that of the electron). On the left-hand sides, the first (second) subscript indicates the polarization mode of the incoming (scattered) photon.
A key feature of this scattering process is that the ordinary mode photons are less likely to scatter (by a factor of three for an isotropic radiation field), and the photon after scattering is three times more likely to be in the extraordinary mode, regardless of the polarization of the incoming photon.  Consequently, if the absorption feature evident in the XMM-Newton observations is due to resonant cyclotron scattering, the radiation that passes through the plasma without scattering will be dominated by O-mode photons.

To illustrate this, let us assume that the count rate at phase zero in the absence of scattering is 0.275~Hz, and the radiation before scattering is unpolarized.  The rate of scattered photons is 0.150~Hz (from the decrease in flux during the Big Dip as observed by IXPE); so, if the rate of scattering of extraordinary and ordinary photons is 0.090~Hz and 0.060~Hz, respectively (to total 0.150~Hz), the unscattered radiation that we observe at phase zero would have $U=0.030$~Hz (dominated by the ordinary mode) and a polarization degree of about $25\%$, as observed.  A modest difference in the scattering probability of $50\%$ in this example is sufficient to account for the observed polarization in the minimum.  Two thirds of the scattered photons emerge in the extraordinary mode and one third in the ordinary mode, so over the three phase bins, where the scattering occurs, 
a net of 0.05 photons per second are scattered into the extraordinary mode (i.e $\textrm{O}\rightarrow\textrm{X}$).  If we assume that these are visible over the five phase bins from 0.18 to 0.46, the average rate is 0.03~Hz, as observed in these phases at the corresponding polarization angle, if the dominant mode does indeed switch from ordinary to extraordinary at about phase 0.1.  The presence of scattered photons in the Big Peak can account for the observed polarization at this phase of the star's rotation.  On the other hand, the Little Peak is not appreciably polarized and is somewhat smaller in amplitude; both these features are expected if the scattered photons do not contribute at this phase.  In principle the plasma loop is simply hidden behind the star during the Little Peak until its footprints appear over the horizon at the beginning of the Big Dip and the loop begins to block our line-of-sight to the emission region.  
Clearly, this scenario requires a complicated geometry for the magnetic field near the surface that does not correspond to the large-scale dipole field of the neutron star. This should in principle rule out an explanation of the observed polarization angle in terms of a simple RVM. The fact that the RVM does indeed provide a reliable explanation to the data is further evidence that the polarization direction is determined at a distance far away from the surface (i.e. the polarization-limited radius), where the field and therefore the polarization vectors are aligned with the global dipole direction.

\section{Spectropolarimetric Modelling}
\label{spectropol_model}

\begin{figure}
\begin{center}
\includegraphics[width=\linewidth]{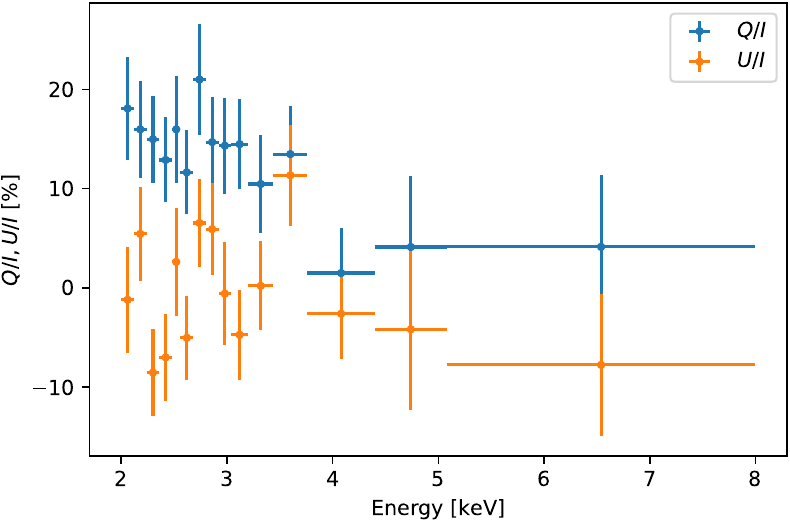}
\caption{Normalized Stokes parameters $Q/I$ and $U/I$ plotted as functions of energy, referred to the frame of the best-fitting RVM with mode switching (dashed curve in Figure~\ref{figure:bestfitRVM}).}
\label{figure:q-rvm}
\end{center}
\end{figure}

We next examine the extent of polarization averaged over the rotational phase as a function of energy.  In order to get a better sense of the polarization properties at the emission, the polarization angles of the photons at each phase are measured relative to the best-fitting RVM (with mode switching, the dashed curve in the third panel of Figure~\ref{figure:bestfitRVM}). The results are reported in Figure~\ref{figure:q-rvm}. 
In particular the values of $U/I$ are consistent with zero which reflects the fact that the polarization states of the photons are conserved as the photons travel out to the polarization-limiting radius \citep{2000MNRAS.311..555H}.  If there were a substantial source of polarized emission outside the polarization-limiting radius, the value of $U/I$ would not necessarily vanish. The component of polarization along and across the dipole axis ($Q/I$) ranges from near $20\%$ at $2$~keV and drops to be essentially consistent with zero above $4$~keV.  This dovetails with the hypothesis that the observed polarization is generated by resonant cyclotron scattering off of protons, 
as the proton cyclotron line lies at lower energies in the middle of the Big Dip where the polarization is strongest.

\begin{figure}
\begin{center}
\includegraphics[width=\linewidth]{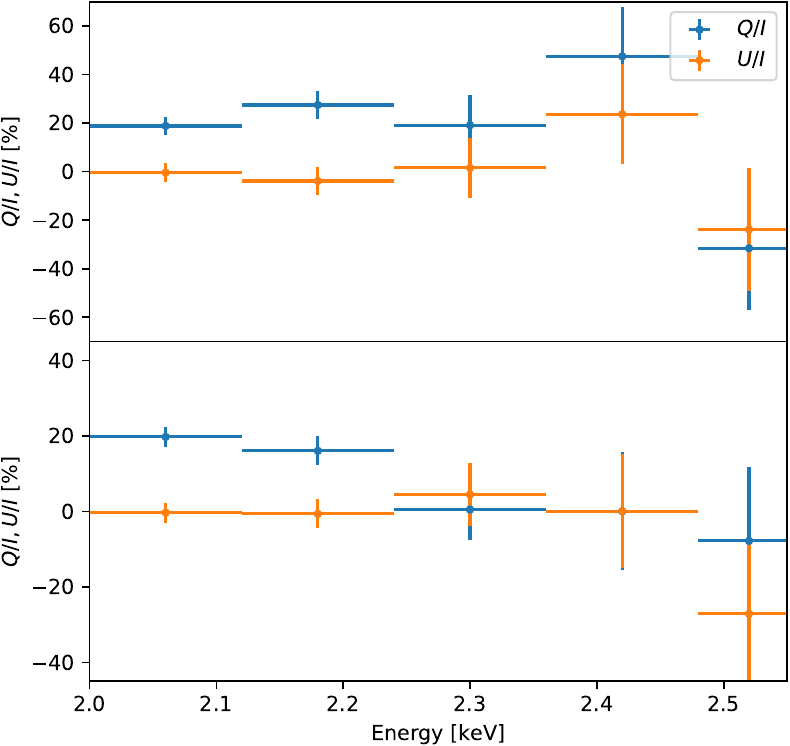}
\caption{Same as in Figure~\ref{figure:q-rvm} but for the Big Dip (upper panel) and the Big Peak (lower panel) phase intervals, restricted to the $2$--$2.6$~keV range. Both $Q$ and $U$ are measured relative to the best-fitting RVM.
}
\label{figure:q-rvm-big}
\end{center}
\end{figure}

We further examine the energy dependence of the polarization as a function of phase.  We focus on just the two regions with large polarized fractions, the Big Dip and the Big Peak and on a relatively narrow range in energy, because the number of photons turns out to be insufficient to reliably determine the polarization above $2.6$~keV in these phase intervals. The results are shown in Figure~\ref{figure:q-rvm-big}.  In both the regions considered, we find that the values of $U/I$ are consistent with zero, as expected.  The polarization degree in the Big Dip (upper panel) is essentially constant across the band, indicating that an equal fraction of photons are scattered as a function of energy. 
Since the spectral feature is broad (see the bottom panel of Figure~\ref{figure:bestfitRVM} and Section~\ref{sec:spectral-feature}), this is not surprising.  On the other hand, the lower panel of Figure~\ref{figure:q-rvm-big} shows that the polarization degree decreases with energy and is consistent with zero above $2.3$~keV.  This feature can also be explained in the scattering picture after we better understand the spectral components as a function of phase (see Section~\ref{sec:broadband}). 

To gain further understanding of the spectral behaviour of the source, we performed the same spectral fitting within each of the phase bins defined in Figure~\ref{fig:folding} and Table~\ref{tab:polarization}.  We also fit the 2023 XMM-Newton EPIC-pn spectra within these bins with and without an absorption feature (the results are reported in Table~\ref{tab:spectral_phasedep}). The fits with and without the feature yield acceptable $\chi^2$ values for the Big Dip and Little Dip, with fits with the spectral feature being preferred.  However, for the Big Peak and Little Peak, the fits without the spectral feature are unacceptable. Interestingly, during the Rise, where, according to Figure~\ref{figure:bestfitRVM}, the line is quickly changing in energy, neither fit is acceptable.  In all of the fits, the energy of the spectral line is about $1$~keV, with a width about $0.2$~keV and a depth $\approx0.15$~keV.  The emission during the Big Peak is typically harder than in the Big Dip.  We found that the fraction of scattered photons during the Big Dip was approximately constant with energy (upper panel of Figure~\ref{figure:q-rvm-big}).  If these photons are preferentially scattered into the extraordinary mode, the contribution of the scattered photons at another phase will be largest where the spectrum of incoming photons is largest. Because the Big Dip is relatively softer than the Big Peak, the relative contribution of the scattered photons to the observed flux will be larger at lower energies, resulting in a decrease of the polarization degree with energy (lower panel of Figure~\ref{figure:q-rvm-big}). 
\begin{table*}
    \centering
    \caption{Results of phase-dependent spectral modelling of the XMM-Newton EPIC-pn data using the {\tt phabs}$\times${\tt (bbodyrad+powerlaw)} and {\tt phabs}$\times${\tt (bbodyrad+powerlaw)}$\times${\tt gabs} decompositions$\,^a$.}
    \label{tab:spectral_phasedep}
    \begin{tabular}{l|cccccccc}
        \hline \hline
\multicolumn{1}{c}{Region} &
\multicolumn{1}{c}{$kT_\textrm{BB}$} & 
\multicolumn{1}{c}{$R_\textrm{BB}$$^b$} &
\multicolumn{1}{c}{$\Gamma_\textrm{PL}$} & 
\multicolumn{1}{c}{$\textrm{Norm}_\textrm{PL}$ at 1~keV} &
\multicolumn{1}{c}{$E_\textrm{abs}$} & 
\multicolumn{1}{c}{$\sigma_\textrm{abs}$} & 
\multicolumn{1}{c}{Depth$_\textrm{abs}$} & 
\multicolumn{1}{c}{$\chi^2/\textrm{dof}$} 
 \\ 
     &       
\multicolumn{1}{c}{[keV]} & 
\multicolumn{1}{c}{[km]} &
  & 
\multicolumn{1}{c}{[$10^{-3}$/s/keV/cm$^{2}$]} &
\multicolumn{1}{c}{[keV]} & 
\multicolumn{1}{c}{[keV]} & 
\multicolumn{1}{c}{[keV]} & 

\\ \hline \hline
    Big Dip     & $ 0.431_{-0.014}^{+0.014}$  &   $ 1.87_{-0.18}^{+0.18}$  &   $4.15_{-0.07}^{+0.08}$  &   $ 39.0_{-2.1}^{+3.0}$  & $1.07_{-0.06}^{+0.08}$ & $0.15_{-0.07}^{+0.08}$ & $0.05_{-0.03}^{+0.05}$ & 75.8 / 78 \\
                &  &  &  &  &  &  &  & \\
                & $ 0.411_{-0.010}^{+0.010}$  &   $ 2.30_{-0.13}^{+0.15}$  &   $3.95_{-0.06}^{+0.06}$  &   $ 29.2_{-0.8}^{+0.8}$  & ---~~~~~~  & ---~~~~~~   & ---~~~~~~  & 80.9 / 81            \\
                &  &  &  &  &  &  &  & \\
                &  &  &  &  &  &  &  & \\
    Rise        & $ 0.401_{-0.026}^{+0.034}$  &   $ 2.49_{-0.62}^{+0.59}$  &   $4.28_{-0.15}^{+0.19}$  &   $64.7_{-8.5}^{+2.1}$  & $0.96_{-0.13}^{+0.08}$ & $0.25_{-0.10}^{+0.16}$ & $0.16_{-0.10}^{+0.36}$ & 73.4 / 63 \\
                &  &  &  &  &  &  &  & \\
                & $0.373_{-0.013}^{+0.013}$  &   $ 3.55_{-0.29}^{+0.35}$  &   $ 3.98_{-0.07}^{+0.08}$  &   $40.8_{-1.7}^{+1.6}$  & ---~~~~~~  &---~~~~~~   & ---~~~~~~  & 77.3 / 66            \\
                &  &  &  &  &  &  &  & \\
                &  &  &  &  &  &  &  & \\
    Big Peak    & $ 0.423_{-0.011}^{+0.011}$  &   $ 2.26_{-0.16}^{+0.17}$  &   $ 3.90_{-0.04}^{+0.04}$  &   $54.0_{-2.0}^{+2.4}$  & $1.06_{-0.03}^{+0.03}$ & $0.14_{-0.04}^{+0.04}$ & $0.05_{-0.02}^{+0.02}$ & 106.1 / 103 \\
                &  &  &  &  &  &  &  & \\
                & $ 0.402_{-0.007}^{+0.007}$  &   $ 2.89_{-0.13}^{+0.14}$  &   $ 3.71_{-0.03}^{+0.03}$  &   $39.7_{-0.8}^{+0.8}$  & ---~~~~~~  &---~~~~~~   & ---~~~~~~  & 122.9 / 106            \\  
                &  &  &  &  &  &  &  & \\
                &  &  &  &  &  &  &  & \\
    Little Dip  & $ 0.476_{-0.037}^{+0.048}$  &   $ 1.00_{-0.25}^{+0.29}$  &   $ 3.94_{-0.08}^{+0.11}$  &   $38.1_{-3.3}^{+6.2}$  & $1.10_{-0.10}^{+0.08}$ & $0.25_{-0.09}^{+0.13}$ & $0.11_{-0.06}^{+0.17}$ & 71.2 / 78 \\
                &  &  &  &  &  &  &  & \\
                & $ 0.435_{-0.018}^{+0.018}$  &   $ 1.55_{-0.14}^{+0.18}$  &   $ 3.74_{-0.05}^{+0.06}$  &   $27.0_{-0.7}^{+0.7}$  & ---~~~~~~  &---~~~~~~   & ---~~~~~~  & 74.3 / 81            \\
                &  &  &  &  &  &  &  & \\
                &  &  &  &  &  &  &  & \\
    Little Peak & $ 0.407_{-0.014}^{+0.015}$  &   $ 2.23_{-0.26}^{+0.26}$  &   $ 3.91_{-0.04}^{+0.06}$  &   $56.0_{-3.1}^{+5.1}$  & $1.09_{-0.05}^{+0.04}$ & $0.19_{-0.05}^{+0.07}$ & $0.08_{-0.03}^{+0.06}$ & 103.5 / 100 \\ 
                &  &  &  &  &  &  &  & \\
                & $ 0.393_{-0.008}^{+0.008}$  &   $ 2.90_{-0.15}^{+0.16}$  &   $ 3.70_{-0.03}^{+0.03}$  &   $39.1_{-0.8}^{+0.8}$  & ---~~~~~~  &---~~~~~~   & ---~~~~~~& 122.3 / 103            \\
\hline  
    \end{tabular}
\begin{list}{}{}
\item[\,\,\,\,] Errors are at $1\sigma$ confidence level.
\item[$^a$] The column density parameter is frozen to that returned by the correspondent best fitting model of the phase-integrated EPIC-pn data (see section \ref{sec:broadband}), i.e. $N_{\rm H}=0.96\times10^{22}\,{\rm cm}^{-2}$ and $1.02\times10^{22}\,{\rm cm}^{-2}$, respectively. 
\item[$^b$] Derived by adopting a $3.2$ kpc distance \citep{kf12,2019A&A...626A..39P}.
\end{list}
\end{table*}
\subsection{The spectral feature}
\label{sec:spectral-feature}

\begin{figure}
    \centering
    \includegraphics[width=\linewidth]{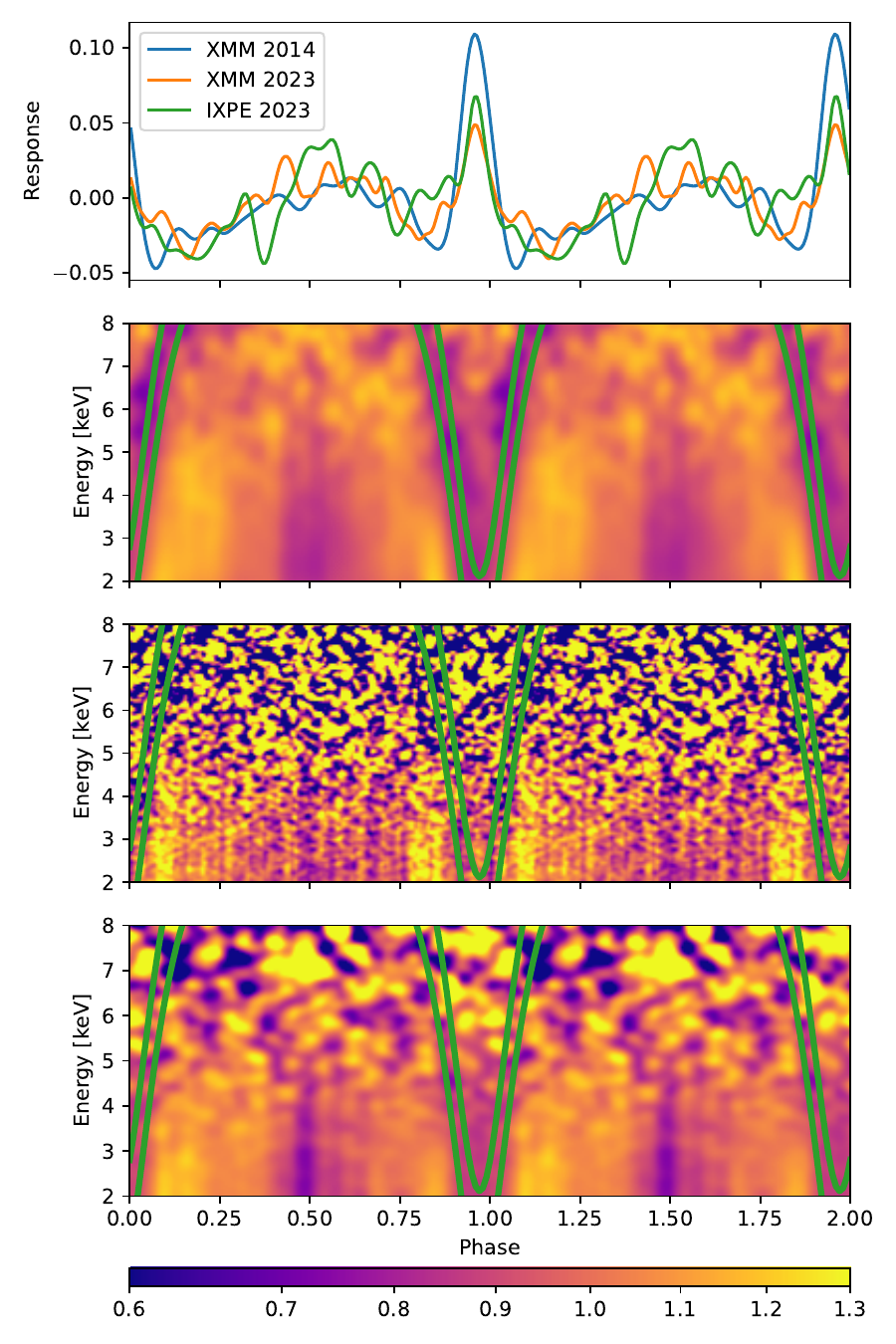}
    \caption{Dynamic Phase- and Energy-Normalized Spectra.  Upper: Response using a matched filter defined by the line detected in the XMM-Newton 2014 EPIC-pn data when applied to the XMM-Newton EPIC-pn data from 2014, EPIC-pn and MOS data from 2023, and the IXPE data in 2023.  The response finds the line feature in all three datasets centered at about the phase 0.97.
    Upper-Middle: the dynamic spectrum from XMM-Newton EPiC-pn in 2014 with the defined feature region from Eq.~\ref{eq:feature} superimposed. 
    Lower-Middle: the dynamic spectrum from XMM-Newton EPIC-pn and MOS in 2023. Lower: the dynamic spectrum from IXPE in 2023.  Both are normalized by the phase-averaged energy spectrum and then the energy-integrated pulse profile \citep[as Fig. 2 of][and Fig.~\ref{figure:bestfitRVM} above]{2019A&A...626A..39P}.  }
    \label{fig:line-region}
\end{figure}

We define the region in energy and phase containing the spectral feature as centred on 
\begin{equation}
E = 12.3~\textrm{keV} -  \frac{11.9~\textrm{keV}}{1+78.4 x^2}\,,
\label{eq:feature}
\end{equation}
where
\begin{equation}
x = \left \{ \frac{\phi}{2\pi} - 0.971 + \frac{1}{2} \right \} - \frac{1}{2}
\label{eq:feature_centre}
\end{equation}
and $\{\}$ denotes the fractional part and $\phi$ is the rotational phase in radians. We take the width of the feature to be $2$~keV.  The particular numerical values in equations~(\ref{eq:feature}) and~(\ref{eq:feature_centre}) were determined by finding the region of width $2$~keV with the smallest mean value of the phase-resolved spectrum normalized by the phase-averaged energy spectrum and then by the energy-integrated pulse profile (the upper middle panel of  Figure~\ref{fig:line-region}).  By design, the mean at each phase is unity.  For the XMM-Newton observation, the standard deviation of the mean over a region with the area delineated in the upper panel is 0.01, whereas the value obtained in the XMM-Newton energy-phase region is 0.86 (fourteen standard deviations below the expected value).  For IXPE in the similarly sized region the standard deviation is 0.005 and the value obtained is 0.88 (twenty-four standard deviations below the expected value), indicating with high confidence that the spectral feature is also present in the recent IXPE observations.  The upper most panel of Figure~\ref{fig:line-region} depicts the response of a filter with the shape of the feature against the three datasets.  In all three the feature is detected with the matched filter.

We can exploit the time and energy-resolution of IXPE to calculate the polarization degree for only the events within the phase and energy domain delineated by equation~(\ref{eq:feature}) within the Big Dip phase interval, $(0.86,0.07)$.  We obtain $Q/I=0.05 \pm 0.06$ and $U/I=0.28\pm0.06$ (measured relative to North, with $\textrm{MDP}_{99}=0.18$).  If we examine the same phase range and exclude the energy range of the feature, we obtain $Q/I=-0.03 \pm 0.04$ and $U/I=0.20\pm0.04$ (with $\textrm{MDP}_{99}=0.11$), demonstrating that the spectral region defined by the feature in the XMM-Newton observation may be  significantly more polarized than the emission at other energies during this phase of the star's rotation.

\section{Conclusions}

Observations of 1E~2259+5586 with IXPE and XMM-Newton support a consistent scenario for the emission from this source. According to our interpretation, the emission is initially only weakly polarized as expected for a condensed surface layer as in 4U~0142+61 \citep{tav+22} and 1RXS~J170849.0--400910 \citep{zane+23}, but during particular phases of the star's rotation the radiation that reaches us passes through a loop of plasma, and protons scatter the radiation in the cyclotron resonance \citep[as in SGR 0418+5729,][]{2013Natur.500..312T}.  As the scattering is preferentially from the ordinary mode into the extraordinary mode, this results in a net polarization in the ordinary mode during the phase where the radiation passes through the loop towards us (the Big Dip), and in the extraordinary mode during the phases where the loop is visible but does not intersect the line-of-sight to the emission region (the Big Peak and Little Dip), leaving us to observe scattered photons.  During the Little Peak, the polarization is weak, presumably because an emission region with its inherently weak polarization is visible, but the plasma loop is hidden from view, so scattered photons do not contribute during this phase.  Further observations of 1E~2259+586 with IXPE could confirm the hint that the polarized flux correlates both in energy and phase with the spectral absorption feature found by \citet{2019A&A...626A..39P}.     

\section*{Acknowledgements}

The Imaging X-ray Polarimetry Explorer (IXPE) is a joint US and Italian mission.  The US contribution is supported by the National Aeronautics and Space Administration (NASA) and led and managed by its Marshall Space Flight Center (MSFC), with industry partner Ball Aerospace (contract NNM15AA18C).  The Italian contribution is supported by the Italian Space Agency (Agenzia Spaziale Italiana, ASI) through contract ASI-OHBI-2022-13-I.0, agreements ASI-INAF-2022-19-HH.0 and ASI-INFN-2017.13-H0, and its Space Science Data Center (SSDC) with agreements ASI-INAF-2022-14-HH.0 and ASI-INFN 2021-43-HH.0, and by the Istituto Nazionale di Astrofisica (INAF) and the Istituto Nazionale di Fisica Nucleare (INFN) in Italy.  This research used data products provided by the IXPE Team (MSFC, SSDC, INAF, and INFN) and distributed with additional software tools by the High-Energy Astrophysics Science Archive Research Center (HEASARC), at NASA Goddard Space Flight Center (GSFC).

This research used data products provided by the IXPE Team (MSFC, SSDC, INAF, and INFN) and distributed with additional software tools by the High-Energy Astrophysics Science Archive Research Center (HEASARC), at NASA Goddard Space Flight Center (GSFC). JH acknowledges support from the Natural Sciences and Engineering Research Council of Canada (NSERC) through a Discovery Grant, the Canadian Space Agency through the co-investigator grant program, and computational resources and services provided by Compute Canada, Advanced Research Computing at the University of British Columbia, and the SciServer science platform (www.sciserver.org). We thank N. Schartel for approving a Target of Opportunity observation with XMM-Newton in the Director’s Discretionary Time (ObsID 0931790), and the XMM-Newton SOC for carrying out the observation.
The work of RTa and RTu is partially supported by the PRIN grant 2022LWPEXW of the Itialian Ministry for University and Reasearch (MUR).
DG-C  acknowledges support from a CNES fellowship grant.  G.L.I. acknowledge financial support from INAF through grant “IAF-Astronomy Fellowships in Italy 2022 - (GOG)”.  I. L. was supported by the NASA Postdoctoral Program at the Marshall Space Flight Center, administered by Oak Ridge Associated Universities under contract with NASA.

\noindent\textit{Software}: XSPEC \citep{Arnaud1996},  Matplotlib \citep{Hunter:2007}, numpy \citep{harris2020array}, heasoft, {\sc ixpeobssim} \citep{BALDINI2022101194}, Astropy \citep{2013A&A...558A..33A,2018AJ....156..123A,2022ApJ...935..167A}
\section*{Data Availability}

    The data used for this analysis are available through HEASARC under IXPE Observation ID 02007899.



\bibliographystyle{mnras}
\bibliography{main}


\section*{List of Affiliations}
\small\textit{\\
$^{1}$ University of British Columbia, Vancouver, BC V6T 1Z4, Canada\\
$^{2}$ Dipartimento di Fisica e Astronomia, Universit\`a degli Studi di Padova, Via Marzolo 8, 35131 Padova, Italy\\
$^{3}$ Mullard Space Science Laboratory, University College London, Holmbury St Mary, Dorking, Surrey RH5 6NT, UK\\
$^{4}$ INAF Osservatorio Astronomico di Roma, Via Frascati 33, 00078 Monte Porzio Catone (RM), Italy\\
$^{5}$ MIT Kavli Institute for Astrophysics and Space Research, Massachusetts Institute of Technology, 77 Massachusetts Avenue, Cambridge, MA 02139, USA\\
$^{6}$ Institut de Recherche en Astrophysique et Plan\'etologie, UPS-OMP, CNRS, CNES, 9 avenue du Colonel Roche, BP 44346 31028, Toulouse CEDEX 4, France\\
$^{7}$ California Institute of Technology, Pasadena, CA 91125, USA\\
$^{8}$ NASA Marshall Space Flight Center, Huntsville, AL 35812, USA\\
$^{9}$ Department of Physics and Astronomy, Louisiana State University, Baton Rouge, LA 70803, USA\\
$^{10}$ Instituto de Astrof\'isica de Andaluc\'ia, CSIC, Glorieta de la Astronom\'ia s/n, 18008 Granada, Spain\\
$^{11}$ Space Science Data Center, Agenzia Spaziale Italiana, Via del Politecnico snc, 00133 Roma, Italy\\
$^{12}$ INAF Osservatorio Astronomico di Cagliari, Via della Scienza 5, 09047 Selargius (CA), Italy\\
$^{13}$ Istituto Nazionale di Fisica Nucleare, Sezione di Pisa, Largo B. Pontecorvo 3, 56127 Pisa, Italy\\
$^{14}$ Dipartimento di Fisica, Universit\`a di Pisa, Largo B. Pontecorvo 3, 56127 Pisa, Italy\\
$^{15}$ Dipartimento di Matematica e Fisica, Universit\`a degli Studi Roma Tre, Via della Vasca Navale 84, 00146 Roma, Italy\\
$^{16}$ Istituto Nazionale di Fisica Nucleare, Sezione di Torino, Via Pietro Giuria 1, 10125 Torino, Italy\\
$^{17}$ Dipartimento di Fisica, Universit\`a degli Studi di Torino, Via Pietro Giuria 1, 10125 Torino, Italy\\
$^{18}$ INAF Osservatorio Astrofisico di Arcetri, Largo Enrico Fermi 5, 50125 Firenze, Italy\\
$^{19}$ Dipartimento di Fisica e Astronomia, Universit\`a degli Studi di Firenze, Via Sansone 1, 50019 Sesto Fiorentino (FI), Italy\\
$^{20}$ Istituto Nazionale di Fisica Nucleare, Sezione di Firenze, Via Sansone 1, 50019 Sesto Fiorentino (FI), Italy\\
$^{21}$ INAF Istituto di Astrofisica e Planetologia Spaziali, Via del Fosso del Cavaliere 100, 00133 Roma, Italy\\
$^{22}$ ASI - Agenzia Spaziale Italiana, Via del Politecnico snc, 00133 Roma, Italy\\
$^{23}$ Science and Technology Institute, Universities Space Research Association, Huntsville, AL 35805, USA\\
$^{24}$ Istituto Nazionale di Fisica Nucleare, Sezione di Roma "Tor Vergata", Via della Ricerca Scientifica 1, 00133 Roma, Italy\\
$^{25}$ Department of Physics and Kavli Institute for Particle Astrophysics and Cosmology, Stanford University, Stanford, California 94305, USA\\
$^{26}$ Institut f\"ur Astronomie und Astrophysik, Universit\"at T\"ubingen, Sand 1, 72076 T\"ubingen, Germany\\
$^{27}$ Astronomical Institute of the Czech Academy of Sciences, Bo\v{c}n\'i II 1401/1, 14100 Praha 4, Czech Republic\\
$^{28}$ RIKEN Cluster for Pioneering Research, 2-1 Hirosawa, Wako, Saitama 351-0198, Japan\\
$^{29}$ NASA Goddard Space Flight Center, Greenbelt, MD 20771, USA\\
$^{30}$ Yamagata University,1-4-12 Kojirakawa-machi, Yamagata-shi 990-8560, Japan\\
$^{31}$ Osaka University, 1-1 Yamadaoka, Suita, Osaka 565-0871, Japan\\
$^{32}$ International Center for Hadron Astrophysics, Chiba University, Chiba 263-8522, Japan\\
$^{33}$ Institute for Astrophysical Research, Boston University, 725 Commonwealth Avenue, Boston, MA 02215, USA\\
$^{34}$ Department of Astrophysics, St. Petersburg State University, Universitetsky pr. 28, Petrodvoretz, 198504 St. Petersburg, Russia\\
$^{35}$ Department of Physics and Astronomy and Space Science Center, University of New Hampshire, Durham, NH 03824, USA\\
$^{36}$ Physics Department and McDonnell Center for the Space Sciences, Washington University in St. Louis, St. Louis, MO 63130, USA\\
$^{37}$ Istituto Nazionale di Fisica Nucleare, Sezione di Napoli, Strada Comunale Cinthia, 80126 Napoli, Italy\\
$^{38}$ Universit\'e de Strasbourg, CNRS, Observatoire Astronomique de Strasbourg, UMR 7550, 67000 Strasbourg, France\\
$^{39}$ Graduate School of Science, Division of Particle and Astrophysical Science, Nagoya University, Furo-cho, Chikusa-ku, Nagoya, Aichi 464-8602, Japan\\
$^{40}$ Hiroshima Astrophysical Science Center, Hiroshima University, 1-3-1 Kagamiyama, Higashi-Hiroshima, Hiroshima 739-8526, Japan\\
$^{41}$ Department of Physics, The University of Hong Kong, Pokfulam, Hong Kong\\
$^{42}$ Department of Astronomy and Astrophysics, Pennsylvania State University, University Park, PA 16802, USA\\
$^{43}$ Universit\'e Grenoble Alpes, CNRS, IPAG, 38000 Grenoble, France\\
$^{44}$ Department of Physics and Astronomy, 20014 University of Turku, Finland\\
$^{45}$ Center for Astrophysics | Harvard \& Smithsonian, 60 Garden St, Cambridge, MA 02138, USA\\
$^{46}$ INAF Osservatorio Astronomico di Brera, Via E. Bianchi 46, 23807 Merate (LC), Italy\\
$^{47}$ Dipartimento di Fisica, Universit\`a degli Studi di Roma "Tor Vergata", Via della Ricerca Scientifica 1, 00133 Roma, Italy\\
$^{48}$ Department of Astronomy, University of Maryland, College Park, Maryland 20742, USA\\
$^{49}$ Anton Pannekoek Institute for Astronomy \& GRAPPA, University of Amsterdam, Science Park 904, 1098 XH Amsterdam, The Netherlands\\
$^{50}$ Guangxi Key Laboratory for Relativistic Astrophysics, School of Physical Science and Technology, Guangxi University, Nanning 530004, China\\
}



\bsp	
\label{lastpage}
\end{document}